\documentclass{article}

\usepackage{arxiv}

\usepackage[utf8]{inputenc} % allow utf-8 input
\usepackage[T1]{fontenc}    % use 8-bit T1 fonts
\usepackage{hyperref}       % hyperlinks
\usepackage{url}            % simple URL typesetting
\usepackage{booktabs}       % professional-quality tables
\usepackage{amsfonts}       % blackboard math symbols
\usepackage{nicefrac}       % compact symbols for 1/2, etc.
\usepackage{microtype}      % microtypography
\usepackage{lipsum}
\usepackage[font=small,labelfont=bf]{caption} 

\usepackage{graphicx,calc}
\usepackage{amsmath}
\usepackage[english]{babel}
\usepackage{eurosym}
\usepackage{stmaryrd}
\usepackage{xcolor}
\usepackage{comment}
\usepackage{easylist}

\usepackage{isomath}
\usepackage{amsmath}
\usepackage{amssymb}
\usepackage{amscd}
\usepackage{amsfonts}

\newcommand{\beq}{\begin{equation}}
\newcommand{\eeq}{\end{equation}}
\newcommand{\beqs}{\begin{eqnarray}}
\newcommand{\eeqs}{\end{eqnarray}}

\usepackage{gensymb}
\usepackage{tabularx}
\usepackage{fancyhdr}
\usepackage{graphics}
\usepackage{verbatim}
\usepackage{subfigure}
\usepackage{xspace}
\usepackage{euscript}
\usepackage{alltt}
\usepackage{boxedminipage}
\usepackage{float}
\usepackage{color}
\usepackage{bbold}
\usepackage{mathtools}
\usepackage{hyperref}
\usepackage[all]{xy}
\usepackage{t1enc}
\usepackage{times,exscale}
\usepackage{graphicx,calc}
\usepackage{mdwlist}
\usepackage{framed}
\usepackage{enumitem}
\usepackage{stmaryrd}
\numberwithin{equation}{section}

\everymath{\displaystyle}
\usepackage{booktabs}
\usepackage{multirow}
\usepackage{graphicx}
\usepackage{setspace}
\usepackage{siunitx}
\usepackage{textcomp}

\newenvironment{sistema}%
{\left\lbrace\begin{array}{@{}l@{}}}%
{\end{array}\right.}

\graphicspath{{IMGnew//}}

\title{Toughening and mechanosensing in bone: a perfectly balanced mechanism based on competing stresses}

\author{
 M. Fraldi\\
  Department of Structures for Engineering and Architecture, University of Napoli Federico II, Italy\\
    \texttt{fraldi@unina.it} \\
  \And
  A. Cutolo \\
  Department of Structures for Engineering and Architecture, University of Napoli Federico II, Italy\\
  \texttt{arsenio.cutolo@unina.it} \\
  \And   
   A. R. Carotenuto \\
  Department of Structures for Engineering and Architecture, University of Napoli Federico II, Italy\\
  \texttt{angelorosario.carotenuto@unina.it} \\
  \And  
  S. Palumbo \\
  Department of Structures for Engineering and Architecture, University of Napoli Federico II, Italy\\
  \texttt{stefania.palumbo@unina.it} \\
  \And
  F. Bosia \\
  Department of Applied Science and Technology, Politecnico di Torino, Italy\\
  \texttt{federico.bosia@polito.it}\\
  \And
  N.M. Pugno\\
  Department of Civil, Environmental and Mechanical Engineering, University of Trento, Italy\\
  School of Engineering and Materials Science, Queen Mary University, London, UK\\
   \texttt{nicola.pugno@unitn.it} 
}

\begin{document}
\maketitle

\begin{abstract}

Bone is a stiff and though, hierarchically organized and continuously evolving material that optimizes its structure across the scales to properly respond to mechanical stimuli, which also govern growth and remodelling processes through a complex cascades of interlaced mechanobiological events. However, a full understanding of the fascinating underlying mechanisms responsible for the cooperation of bone toughness and biological functions, with important implications in bone ageing, osteoporosis and post-trauma repairing processes, has yet to be achieved. In particular, how micro-damage nucleation --which is necessary for tissue remodelling-- does not evolve into catastrophic failure in such a stiff material, still remains a partial enigma, given that the presence of cement lines, interfaces and sacrificial elements, which dissipate energy and deviate cracks, alone do not provide a definitive answer to the question. To help solve this challenging problem, here we bring to light a novel stress-based bone toughening mechanism, calling into play the nearly-symmetrical, chiral and hierarchical architecture of the osteon, in which adjacent lamellae are arranged in clockwise and counter-clockwise manners as a result of the different orientation of their components, i.e. collagen fibrils and carbonated hydroxyapatite crystallites. Somewhat counter-intuitively, we demonstrate that this arrangement simultaneously gives rise to stress states that are alternating in sign along the osteon radius and to localized stress amplification phenomena, both in the tensile and compressive regimes. This unveils a previously unforeseen synergistic mechanism allowing micro-damage accumulation without propagating cracks, which is kindled by the contrast between crack-opening due to tensile hoop stresses (required for bone remodelling) and crack-stopping due to compressive stresses at the crack tips in adjacent lamellae. This allows to seal the crack ends and thus confer toughness to bone well beyond the level predicted by current models. Furthermore, shear stresses alternating in sign occur at the lamellar interfaces, contributing with fluid flow to mechanically stimulate the osteocytes in the lacunae and thus amplify the signalling for osteoblasts and osteoclasts. These results, obtained through original exact solutions based on the theory of anisotropic elasticity and confirmed by both Finite Element fracture analyses and experimental tests on 3D-printed osteon prototypes, contribute with another piece in the puzzle to making the rational biophysical picture of bone mechanobiology complete.
\\

\end{abstract}

% keywords can be removed

\section{Introduction}
\label{Intro}
\begin{figure}[htbp]
\centering
\includegraphics[width=1\textwidth]{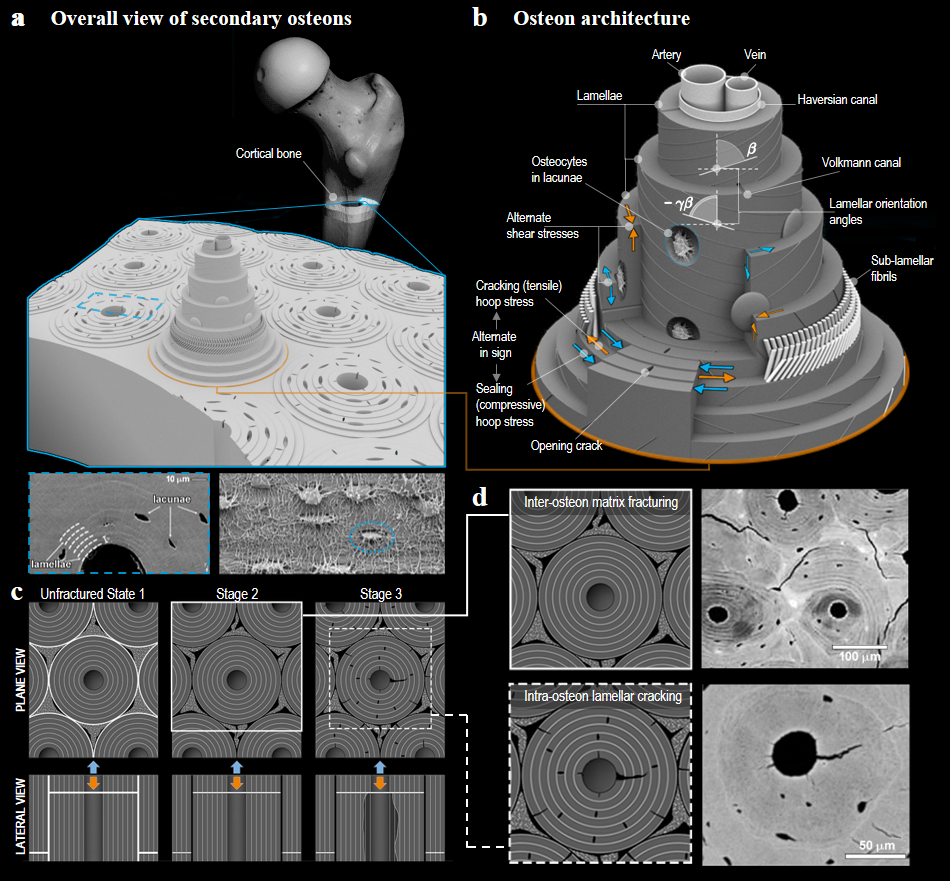}
\caption{Organization, structure and (stress-based) fracturing mechanisms in secondary osteons of cortical bone: \textbf{a)} reconstruction of the three-dimensional arrangement of secondary osteons in an ideal section of cortical bone (in the insets below, the Scanning Electron Microscopy (SEM) images highlighting the lamellae and lacunae system (bottom-left) and a detail of osteocytes hosted in the lacunae at the lamellar interfaces (bottom-right), (SEM adapted from \cite{carnelli,martin-book}); \textbf{b)} architecture of the secondary osteon with emphasis on  the overall helicoidal orientation $\beta$ of the lamellae (which confers monoclinic-trigonal anisotropy to the osteon), the slight wrap angle discrepancy between adjacent lamellae (described by $\gamma$), the sub-lamellar fibril microstructure and the intra-osteon micro-crack opening along the radial direction (in orange and cyan, tensile or compressive hoop stresses, as well as associated shear stresses, which are alternating in sign --in both the cases of tensile and compressive applied loads at the osteon level-- at the basis of the new stress-induced toughening mechanism discussed here); \textbf{c)} schematic of cross and lateral view sections of the osteons to illustrate how damage and micro-fractures nucleate and diffuse in cortical bone due to arbitrary positive or negative applied axial loads (cyan and orange arrows), by first involving the cement lines and progressing in the inter-osteon matrix (stage 2) and then invading osteons, with cracks propagating in radial directions, but somehow surprisingly stopping despite bone being such a stiff material (stage 3); \textbf{d)} comparisons of the outlined stages 2 and 3 and the actual evidence observed experimentally (SEM images adapted from \cite{Milovanovic}).}
\label{fig.1}
\end{figure}

Bone is a unique material that plays a crucial role in key vertebrates' life functions, such as protection of internal organs, definition of morphology, structural support for the whole body, movement and locomotion \cite{Cowin1981,currey2006bones}. This hierarchically organized and continuously evolving mineralized tissue is capable of self-repairing, renewing and adapting its architecture across various spatial scales to maximize stiffness, toughness (the ability to absorb mechanical energy up to the point of failure) and strength in response to mechanical stimuli \cite{cowin1983m,cowin1976,hegedus1976,cowin2003ad,fratzl2007,zimmermann2015,gao2016}. All these features make it one of the smartest known biological materials \cite{Launey2010,Martin2015}.\\
At the macroscopic level, bone essentially appears in both the forms of compact and spongy-like tissues that bear normal and shear mechanical stresses in axial, bending and torque regimes induced by external forces \cite{Cowin2007,zysset1999}. In particular, cortical bone is mainly designed to absorb the highest stresses and to redirect them to optimize the material response. Instead, the trabecular regions --found for instance in the epiphysis of long bones or in innermost sections of the vertebral bodies-- provide, according to the sites, the transfer of loads from joints to distal compact bone districts \cite{oftadeh2015biomechanics}, distributing forces over large articular surfaces, minimizing pressures and maximizing stiffness/weight ratios through growth and remodelling processes that reorganize the bone density distribution and the anisotropic porous skeleton over time \cite{Cowin1986w,cowin2012mixture,cowin1976}. At the microscale, the fundamental units of the cortical bone are (secondary) osteons (see Fig. \ref{fig.1}). These are hollow cylinders with diameters of the order of hundreds of microns (typically from $200$ $\mu m$ to $500$ $\mu m$), made of several concentric layers, named lamellae, which wrap around a central Haversian canal that, together with a transversal network of Volkmann's channels and a diffuse system of canaliculi, hosts nerve fibres and blood vessels to supply nutrients throughout the tissue \cite{Gupta2006a,Weiner1999,cowin2015flow,cardoso2013ad}. In turn, each lamella in the osteon comprises helically arranged collagen fibrils forming angles from about 10° to 60° with respect to the osteon axis, in alternate clockwise and counter-clockwise directions in successive lamellae \cite{Ascenzi1968,Ascenzi2003, Ascenzi2006,Wagermaier2006a,Kazanci2006, Giraud-Guille1988,Schrof2014}, this resulting into a highly anisotropic structural organization, whose complexity is further enriched by the presence of interfaces and material discontinuities.
Most importantly, the osteons contain the main bone cells, namely osteoblasts, osteoclasts and osteocytes, which coordinate their activity to form temporary anatomic structures --the Basic Multicellular Units (BMUs)-- designated to ensure bone integrity and homeostasis through the continuous resorption and deposition of tissue, a process known as bone remodelling \cite{cowin1976,hegedus1976,taber2020}. This is governed by a cascade of mechanical and biochemical signals, occurring across the scales, that develops through a complex network of feedback mechanisms regulating the cells turnover, which are actually not yet completely understood \cite{cowinmicrog}. However, although many questions remain unanswered in the literature and how exactly bone cells sense mechanical loads and then orchestrate their activity is still an open issue, it has been ascertained that micro-damaging, resulting from everyday loadings at the osteon-matrix interfaces and within osteons, triggers the BMU cycle, the level of damage somehow functioning as a stimulus to locally increase mineral content through bone growth and repairing processes \cite{Taylor,Herman2010,cardoso2009ost}. It is in particular claimed that the stress-induced opening of micro-cracks could induce fluid flows, thus promoting nutrient concentration that in turn would lead to intensify bone remodelling activities involving the dismantling of mineralized bone by osteoclasts and the new bone deposition by osteoblasts. The coordination of this process seems to be driven by the osteocytes \cite{klein2013osteocytes,Klein-Nulend,Ericksen,Taylor2,lewis2017osteocyte} --osteoblasts surrounded by the growing bone matrix, which, as the material calcifies, remains trapped in the inter-lamellar spaces, forming lacunae (see Fig. \ref{fig.1})-- that could perform the mechano-sensing function of transducing the interstitial flow and shear micro-strains into amplified biochemical signals, thus contributing to recruit osteoblasts and osteoclasts \cite{cowin2015flow,lewis2017osteocyte,CowinPNAS,Verbruggen2014,wu2013matrix}. This mechanism, at the basis of physiological life processes as well as of osteoporotic disorders, is recognized as one of the most fascinating and effective examples in nature of synergy among biological, chemical and mechanical signals, but a deep understanding of these events has yet to be reached. In particular,  the crucial point of how nucleation of micro-damage does not evolve into catastrophic fracture in such a stiff material as bone is still unclear.\\
To this day, the most substantiated hypothesis to explain how cracks slow down or even stop in bone is essentially related to material anisotropy and to the hierarchical architecture of osteons \cite{Martin2015,wagermaier2015fragility,peterlik2006brittle}. According to this conjecture, after nucleating, cracks would be halted while travelling in the radial direction of an osteon, perceiving the alternate arrangement of the wrapped fibres as an apparent material heterogeneity, first decelerating at the lamellar interfaces, then deviating to propagate circumferentially, simultaneously dissipating energy by both breaking inter-lamellar sacrificial micro-bridge elements and encountering a number of micro-voids and discontinuities. All of this confers special toughness to bone \cite{Nalla2003,Ritchie2021,Ritchie2011,Moyle1984, Libonati,Kendall1975}, including mineralization processes at the fibrils level that determine increasing mechanical properties \cite{ping2022mineralization}. Although some theoretical models and direct observations obtained through imaging techniques have demonstrated that these mechanisms play a role in limiting crack growth, rule of thumb calculations to estimate stress intensity factors at the crack tips and more accurate numerical analyses simulating crack propagation at the osteon level all highlight that multi-scale bone heterogeneity is not \textit{per se} sufficient to arrest cracks or hinder fracture \cite{Ritchie2011}. Moreover, the repairing activity carried out by BMUs, which contributes to both the removal of damaged areas in healthy bone and the preservation of its structural integrity, can neither explain how bone prevents brittle and fatigue failure, since healing processes evidently occur at characteristic times orders of magnitudes greater than the crack speed in a material as stiff as bone. A confirmation that some essential elements are still missing is provided by the fact that realistic static and cyclic loads applied \textit{in vivo} and on bone samples show that cracks do not propagate across all the lamellae, despite both stress analyses performed on detailed (orthotropic) elastic models of osteons under the same loads and experimental findings from composite cylinders with stiffness and strength close to those of bone would predict crack advance or even complete fracture \cite{Libonati, Sabet}. Furthermore, results from these theoretical models suggest that tensile hoop stresses in the lamellae, responsible for micro-crack opening in the radial direction, would occur only in case of compressive loads acting along the osteon axes, while experimental observations highlight lenticular radial micro-cracks generated diffusely in osteons also solicited by tensile loads \cite{Yeni2003,wolfram2016}.\\
On the basis of all these considerations, since the mechanobiology of bone remodelling is triggered by sign-independent (both tensile and compressive) forces acting at the osteon level and is focused on the confinement of micro-crack opening to avoid catastrophic fracture, it must be inferred that some additional, as yet undiscovered and possibly counter-intuitive mechanism should cooperate with (or exploit) bone anisotropy and hierarchical features, in order to guarantee the perfect balance between damage nucleation and crack stopping. To shed light on these still unexplained aspects, this work elucidates an unprecedented crack-arresting and toughening mechanism in bone, by focusing on the mechanical problem of an osteon modelled as a multi-layer hollow cylinder in which each lamella behaves as a monoclinic rather than a simpler orthotropic material, which more faithfully reflects its micro-structure composed of transversely isotropic grouped fibres wound helicoidally, as illustrated in Fig. \ref{fig.1}. Hence, by means of both closed-form elastic solutions and accurate Finite Element (FE) numerical simulations, it is shown how the osteon anisotropy, its hierarchical organization and even some slight asymmetry cooperate to create an unforeseen mechanism of crack-arresting and bone toughening, mediated by hoop and anti-plane shear stresses that alternate their sign in adjacent lamellae and amplify their magnitude at the sub-lamellar scale (see Fig. \ref{fig.1}\hyperref[{fig.1}]{b}). In particular, we prove theoretically that alternate hoop stresses, indifferently due to tensile or compressive axial loads at the osteon level, compete in nucleating and stopping cracks, with tensed cracked lamellae "sealed" upstream and downstream by hoop compressions. Furthermore, we show that shear stresses also are called into play by the chiral osteon architecture, contributing to stimulate osteocytes and so cooperating with the strain amplification phenomena induced by fluid flow shear stresses at the same level \cite{CowinPNAS,weinbaum1994,cowin1998amp,you2001,ICTAM2021}. Finally, we demonstrate that both discrepancy in lamellar wrapping angles and sub-lamellar fibrils arrangement significantly contribute to optimize the discovered new bone toughening mechanism and mechanosensing.

\section{Chirality, deviation from symmetry, hierarchy and alternate stresses in bone mechanobiology}
\label{elastic.section}
From the mechanical point of view, a representative volume element of compact bone can be considered as a composite material made of cylindrical strand structures --the secondary osteons-- mainly oriented in the direction of the principal strains \cite{Lanyon1979} and aligned in long bones parallel to the diaphysis axis, embedded in a stiffer and more brittle matrix, to which they are connected through a thin, highly mineralized interface, known as cement line. As shown by experimental observations and mechanical tests, this implies that, under quasi-static or low-rate cyclic loads, damaging takes place hierarchically \cite{Behiri1984,nalla2005,ZIMMERMANN2014,Wang1997,Chan2010}. The matrix is the first to be affected by micro-cracks, which open and evolve as a function of the magnitude of the exerted forces, and subsequently involve the cement lines.
These in turn behave as weak links \cite{Cowin1981,Nalla2003} that dissipate energy and deviate the crack advancement transversally, by progressively determining a condition in which a large part of the outermost cylindrical surface detaches from the surrounding matrix. This segregation leaves the osteon isolated and only constrained, at its upper and lower bases, to other osteons located above and below it (see scheme of the damaging stages in Fig. \ref{fig.1}\hyperref[{fig.1}]{c-d}) \cite{Martin2015, Krajcinovic1987,BruceMartin1982}. At this stage, compressive or tensile axial forces exerted on cortical bone at the osteon level are then transferred to each osteon body through its bases in terms of imposed displacements, which induce overall axial contraction or elongation. Although the lateral connection with the matrix has now been largely lost, twisting of the osteon, which one would expect from the geometrically unbalanced helical arrangements of the lamellae, remains instead essentially locked by the constrained bases, and this generates reactive torsional shear stresses \cite{lakes1981}. At this juncture, if the applied forces grow, intra-lamellar micro-cracks start to nucleate, then to propagate along the radial direction, finally remaining confined to small regions as a result of a crack-stopping phenomenon that is yet to be fully understood (Fig. \ref{fig.1}\hyperref[{fig.1}]).
\subsection{The osteon as a composite monoclinic elastic hollow cylinder}
In order to model this last stage and try to explain the intra-osteon crack-arresting mechanism, we can neglect poroelastic effects associated to the interstitial fluid flow \cite{cowinporo,cowin2011fabric,cardoso2012role}, limiting at first the mechanical analysis to physiological quasi-static loads, just before damaging and fracturing events take place. To determine the stress state that leads to micro-crack opening, an isolated osteon can be seen as an inhomogeneous, anisotropic and linearly elastic hollow cylinder with traction-free conditions at the outermost cylindrical surface and negligible pressures in the Haversian canal, with the end bases subjected to prescribed displacements in the form of imposed axial contraction/dilation and impeded twisting. In particular, by looking at the hierarchical organization of a secondary osteon, one can recognize a degree of anisotropy that can be correctly described by an elastic Functionally Graded Material Cylinder (FGMC) model \cite{Fraldi2007, Cutolo} made of $n$ perfectly bounded hollow cylindrical phases --representing the lamellae-- comprising helically wrapped and alternate (clockwise and counter-clockwise) families of fibres, which in the case at hand account for mineralized collagen fibrils and carbonated hydroxyapatite crystallites appearing at the sub-lamellar level.\\
Adopting a cylindrical reference system $\left\{r,\varphi,z\right\}$, with the origin placed at the basis of the whole FGMC model (say at the center of the Harvesian canal), the above described micro-structure suggests that the osteon exhibits one plane of material symmetry, say $\varphi-z$, as shown in Fig. \ref{fig.SigGamma2Lam}\hyperref[{fig.SigGamma2Lam}]{a}, which allows to assume cylindrically \textit{monoclinic} anisotropy at the osteon level \cite{Fraldi2002,cowin2013b,cowin1987,zysset1995ft}. This is here retrieved by considering a transverse isotropy in a local helical reference system fixed with respect to the plane orthogonal to the axis of the fibres and then transforming all the quantities of interest in the global cylindrical coordinate system (the mathematical details are reported in Supplementary Information, SI). In this way, the mean helical arrangement of the collagen fibrils and their characteristic alternating wrapping angles in adjacent lamellae --which in previous literature models were neglected by reducing the osteon to a homogenized orthotropic composite cylinder-- have been explicitly taken into account through the monoclinic anisotropy of each lamella ruled by the corresponding wrapping angle. Hence, the generalized Hooke's stress-strain relation $\boldsymbol{\sigma}=\textbf{C}:\boldsymbol{\varepsilon}$ for each lamella $\ell$ will depend on five parameters characterising the transversely isotropic grouped fibres in their own helical reference system and on the layer-specific wrapping angle $\vartheta$, which governs the passage to the cylindrical reference frame in which each lamella, as well as the whole osteon, is monoclinic. In this way, for $\ell=1,...,n$, the stress can be written as $\sigma_{ij}^{(\ell)}=C_{ijhk}^{(\ell)}(E,\nu,\nu_t,\alpha,\eta,\vartheta^{(\ell)})\,\varepsilon_{hk}^{(\ell)}$, while the fourth-rank elasticity tensor is $C_{ijhk}^{(\ell)}=
Q_{im}Q_{jn}Q_{hp}Q_{kq}C_{mnpq}^{hel}(E,\nu,\nu_t,\alpha,\eta)$, where $Q_{mn}(\vartheta^{(\ell)})$ are the components of the orthogonal rotation matrix (explicit transport formulas are recalled in SI), $E$ and $\nu$ are the Young modulus and the Poisson ratio, respectively, in the isotropy plane of the fibres' cross-section, with the Lamé modulus given by $2G=E/(1+\nu)$, and $\alpha=E_t/E$ and $\eta=G_t/G$ are anisotropy coefficients involving in-plane and out-of-plane (denoted by the subscript $t$) moduli referred to the fibres' own helical system (see Table \ref{table} for the numerical values of interest for bone). Equilibrium requires in each lamella that $\text{div} \boldsymbol{\sigma}=\textbf{0}$, while continuity of stresses and displacements at the lamellae interfaces, identified by the radii $r=R^{(\ell)}$, imposes that $\sigma_{rj}^{(\ell)}=\sigma_{rj}^{(\ell+1)}$ and $u_j^{(\ell)}=u_j^{(\ell+1)}$ for $\ell=1,...,n-1$. Additionally, the above mentioned boundary conditions provide vanishing lateral pressures (that is $\sigma_{rr}^{(1)}|_{r=R^{(0)}}=\sigma_{rr}^{(n)}|_{r=R^{(n)}}=0$), prescribed vertical displacements, which can be assigned in terms of applied uniform strain $\epsilon_0\neq0$, and a null twisting angle $\phi_0=0$ at the cylinder bases. Under these conditions, the system of partial differential equations governing the anisotropic elastic problem can be solved analytically, by following a strategy already used by some of the authors \cite{Fraldi2007, Cutolo}, which allows to reduce the problem to a linear algebraic one, then solving it in closed-form for an arbitrary number of lamellae and any set of geometrical and mechanical parameters characterizing the cylindrically monoclinic phases, as explicitly reported in SI.
\subsection{Insights into actual stress states from the simplest two-lamella osteon model}
\begin{figure}[ht]
\centering
\includegraphics[width=\textwidth]{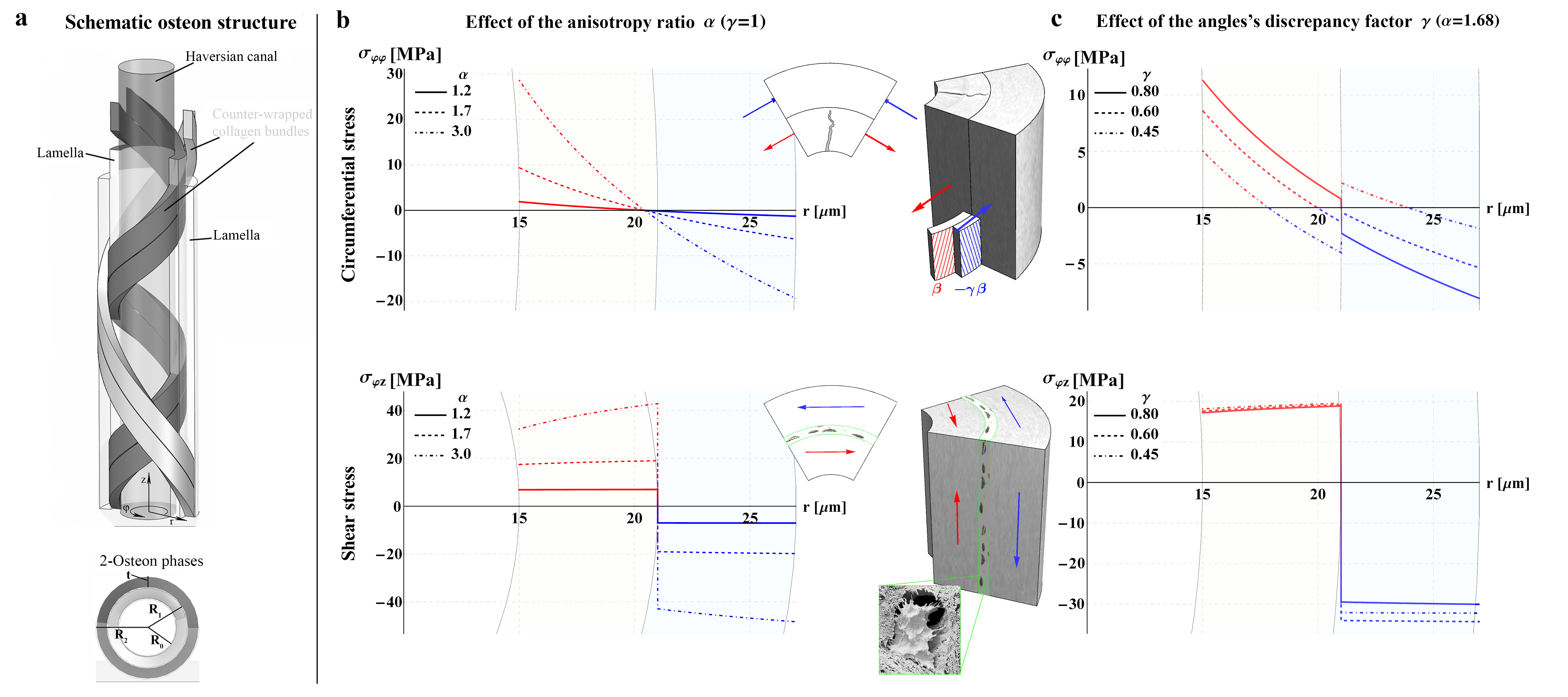}
\caption{\textbf{a)} Schematic structure of a 2-phase osteon. The actual lamellae have a sub-lamellar system comprising many collagen fibrils that are helically arranged in different ways:  these orientations are averaged over each lamella and the mean angles set as $\vartheta^{(1)}=\beta$ and $\vartheta^{(2)}=-\gamma\beta$ respectively, the discrepancy angle $\gamma$ accounting for the observed counter-wrapping imperfection between adjacent lamellae. \textbf{b)} and \textbf{c)} Hoop ($\sigma_{\varphi \varphi}$) and shear ($\sigma_{\varphi z}$) stresses arising in the simplified 2-phase osteon structure subject to a prescribed contraction in the $z$ direction of $1\%$ and locked twisting at its bases, with $\beta=45^{\circ}$. On the left (\textbf{b}) hoop and shear stresses are plotted along the osteon radius by varying the sole anisotropy ratio $\alpha$, in case of perfectly symmetrical counter-wrapping ($\gamma=1$) while on the right (\textbf{c}) the same stresses are plotted when the realistic value of $\alpha=1.68$ \cite{yoon2008} is kept fixed and different possible discrepancy angles are considered ($\gamma=0.45, 0.6, 0.8$). Results, obtained by means of the computational software Mathematica\textsuperscript{\textregistered}\cite{math}, show how counter-intuitive circumferential stresses that are alternate in sign arise along the osteon radius, the tensile components (in red) creating the premises for crack opening and the compressive stresses (in blue) determining a sealing effect that favours the crack arrest. Shear stress also appear at the lamellar interface with alternate signs, thus contributing to stimulate osteocytes in the lacunae and to activate the cascade of signals for the bone mechanotransduction. (The model assumes equal thickness of the lamellae of $6\mu m$ and the Haversian canal radius of $15\mu m$. The other constitutive parameters are those reported in Table \ref{table}).}
\label{fig.SigGamma2Lam}
\end{figure}

Referring the reader to the SI for the detailed analytical solutions, it is instructive to first examine the simpler case of an osteon made of only two cylindrically monoclinic lamellae, each characterized by the same set of five parameters $(E,\nu,\nu_t,\alpha,\eta)$ that are the elastic coefficients and anisotropic ratios of a transversely isotropic material in the helical reference frames defined by $\vartheta^{(1)}=\beta$ and $\vartheta^{(2)}=-\gamma\beta$, which correspond to the (mean) counter-wrapped angles formed by the fibres in the two lamellae with respect to the osteon axis $z$. $\gamma \in \left(0,1\right)$ represents the angle discrepancy factor accounting for the experimentally observed \cite{Vercher-Martnez2015} mismatch (or deviation from symmetry) in microstructural orientation of adjacent lamellae (see Fig. \ref{fig.1}). Therefore, taking numerical values from the literature (see Table \ref{table}) and calculating the analytical solutions in SI for $n=2$, $\beta=45^{\circ}$ and an imposed compressive axial strain of the order of $1\%$, we performed sensitivity analyses by plotting the hoop stresses $\sigma_{\varphi \varphi}$ and the shear stresses $\sigma_{\varphi z}$ along the osteon radius, varying the anisotropy ratio $\alpha$ and the angle discrepancy factor $\gamma$. As shown in Fig. \ref{fig.SigGamma2Lam}, the first remarkable result is that the lamellae experience tensile and compressive circumferential stresses that are unexpectedly alternate in sign\footnote{In fact, with the prescribed boundary conditions for the inner and outer cylindrical surfaces and assuming vanishing in plane shear stresses $\sigma_{r \varphi}$, equilibrium along the radial direction of an infinitesimal circular sector of the osteon implies that, if hoop stresses appear, their integration over the whole thickness should be zero, and hence $\sigma_{\varphi \varphi}$ must change sign along the radius.}, whose peaks grow significantly as the anisotropy ratio increases (Fig. \ref{fig.SigGamma2Lam}\hyperref[fig.SigGamma2Lam]{b-top}). The stress average values over each single lamella also increase, even jumping at the interface as the perfect symmetry of wrapping angles is lost, i.e. as $\gamma<1$, consistently with experimental observations \cite{Vercher-Martnez2015}(see Fig. \ref{fig.SigGamma2Lam}\hyperref[fig.SigGamma2Lam]{c-top}). It is worth highlighting that both average hoop stresses $\sigma _{\varphi \varphi }$ and anti-plane shear stresses $\sigma _{\varphi z}$ have magnitudes comparable with the main axial stresses $\sigma_{zz}$ directly called into play by the applied displacements, while the radial stresses $\sigma_{rr}$ are at least one order of magnitude smaller than the others and the in-plane shear stresses $\sigma _{r \varphi}$ are uniformly vanishing (see SI).
\\
These first results provide a number of relevant insights into how bone, through its hierarchical organization, does not simply use its microstructure to slow down cracks by dissipating energy, but exploits the unique chiral architecture of the osteons and even some apparent geometrical "imperfections" to activate an extremely sophisticated stress-based crack-stopping mechanism. In fact, even at this simplified modelling level, we can trace the main strategies that bone adopts to accommodate two competing demands, i.e. damaging and toughness. First, the theoretical outcomes show that the helical and symmetrical ($\gamma=1$) counter-wrapping of the lamellae on their own creates a hoop stress competitive mechanism in which, as tensile stresses prepare the ground for crack nucleation and opening in the radial direction, compressive stresses simultaneously counter any attempt of cracks to propagate in adjacent regions, generating sealing circumferential forces that halt the crack by impeding the radial diffusion of tensile regimes. This key mechanism is thus significantly amplified, in terms of difference between average hoop compression and tension $\Delta\,\overline{\sigma}_{\varphi\,\varphi}$, by the anisotropy ratio, which in healthy bone osteons can be estimated as $\alpha\approx1.7$ \cite{yoon2008} or greater, as highlighted in Fig. \ref{fig.SigGamma2Lam}\hyperref[{fig.SigGamma2Lam}]{b-top}. In particular, the following helpful rule of thumb can be derived from exact solutions: $\Delta\,\overline{\sigma}_{\varphi\,\varphi} \approx E\,\epsilon_0(\alpha -1) \tau  \sin ^2(\beta) (\cos (2 \beta )+1)/2$, where $\tau$ is a ratio between the thickness of the lamellae (which can be set initially to $1$) and, for the sake of simplicity, some further simplifying assumptions were introduced (see SI). From this equation, one can appreciate that this dimensionless stress gap scales (almost) linearly with $(\alpha-1)$, according to the requirements imposed by the theory of elasticity for the Poisson ratio. From the same expression, it also emerges that chirality and anisotropy cooperate in the basic model of an osteon in making this stress difference effective (e.g. non zero), since the counter-wrapping angle $\beta$ provides a contribution if and only if the material is anisotropic (i.e. $\alpha>1$). The contribution of the anisotropy also becomes vanishing if the chirality is lost, for instance if $\beta=0$.
Furthermore, the mechanism of competing stresses is additionally enhanced by a deviation from perfect symmetry \cite{fraldi2014sef} in lamellar counter-wrapping angles, here represented by the discrepancy factor $\gamma$. Indeed, as highlighted in Fig. \ref{fig.SigGamma2Lam}\hyperref[{fig.SigGamma2Lam}]{c-top}, the mismatch $\Delta\,\beta=(\gamma-1)\,\beta$ between the wrapping angles, once again counter-intuitively, is directly responsible for modifying the almost continuous profile of the hoop stress along the radius, by making it steeply discontinuous at the lamellae interface and in this way amplifying the sealing effect mentioned above as a consequence of an abrupt change of sign of hoop stresses. In particular, rough estimations, still based on analytical solutions (see SI), allow to find at the first order a direct proportionality between the angle mismatch $\Delta\,\beta$ and the hoop stress jump at the interface, say $\llbracket\sigma_{\varphi\varphi}\rrbracket$, which is $\llbracket\sigma_{\varphi\varphi}\rrbracket \approx E\,\epsilon_0(\alpha -1) \Delta\,\beta \sin(2\beta) [(1-\nu)(1+2\nu)\cos (2\beta )+\nu]/(1-\nu^2)$, as confirmed by the sensitivity analyses reported in Fig. \ref{fig.SigGamma2Lam}\hyperref[{fig.SigGamma2Lam}]{c-top} with $\gamma$ varying from $0.45$ to $1$.
Another key result is that chirality, anisotropy and asymmetry also cooperate to give nonzero anti-plane shear stresses $\sigma_{\varphi z}$, also alternate in sign (see Fig. \ref{fig.SigGamma2Lam}\hyperref[{fig.SigGamma2Lam}]{ bottom}), which do not participate to stop crack propagation but directly act to significantly improve the effectiveness of the mechanobiology of the bone at the osteon level. These shear stresses, absent in orthotropic models and therefore neglected in all the previous approaches, directly stimulate the osteocytes in the lacunae, excavated at the interface between lamellae, contributing with the fluid flow to amplify the shear signals that have been demonstrated to be crucial to stimulate the BMU cell activities \cite{CowinPNAS}. Finally, it is worth underlining that both the alternating hoop and shear stresses --and thus the associated above described mechanisms of crack-stopping and osteocyte shear stimulation-- would occur, in an inverted mode, if elongations were applied to the osteon bases instead of contractions. This means that the discussed mechanisms are independent from the sign of applied forces and can thus help to explain how bone can also be capable of working under cyclic bending loads, as confirmed by experimental evidence.    
\begin{table} [hbt]
\centering 
\begin{tabular}{ccc}
\toprule
\textbf{Parameter} & \textbf{Value} \\
\midrule
Fiber Young modulus in the isotropy plane: $E$ & $16 $ GPa \\
Fiber anisotropy ratio: $\alpha=E_t/E$ & $1.68$ \\
Fiber Poisson ratio in the isotropy plane: $\nu$ &  $0.31$ \\
Fiber Poisson ratio in the anisotropy planes: $\nu_t$ & $0.26$\\
Fiber shear moduli ratio $\eta=G_t/G$ & $1$ \\
%Haversian canal radius & $15$ $\mu$m \\
%Lamellar phases' thickness & $6$ $\mu$m \\
%Sub-lamellar phases' thickness & $1$ $\mu$m \\
\bottomrule
\end{tabular}
\caption{Values adopted for the constitutive parameters of the osteon models presented in the present work, based on consolidated literature data \cite{CowinPNAS,Vercher-Martnez2015, Gupta2006a,yoon2008}.}
\label{table}
\end{table} 

\subsection{Alternate crack-opening and crack-halting hoop stresses and osteocyte stimulation in actual osteons}
\label{numeric1}
\begin{figure}[htbp]
\centering
\includegraphics[width=\textwidth]{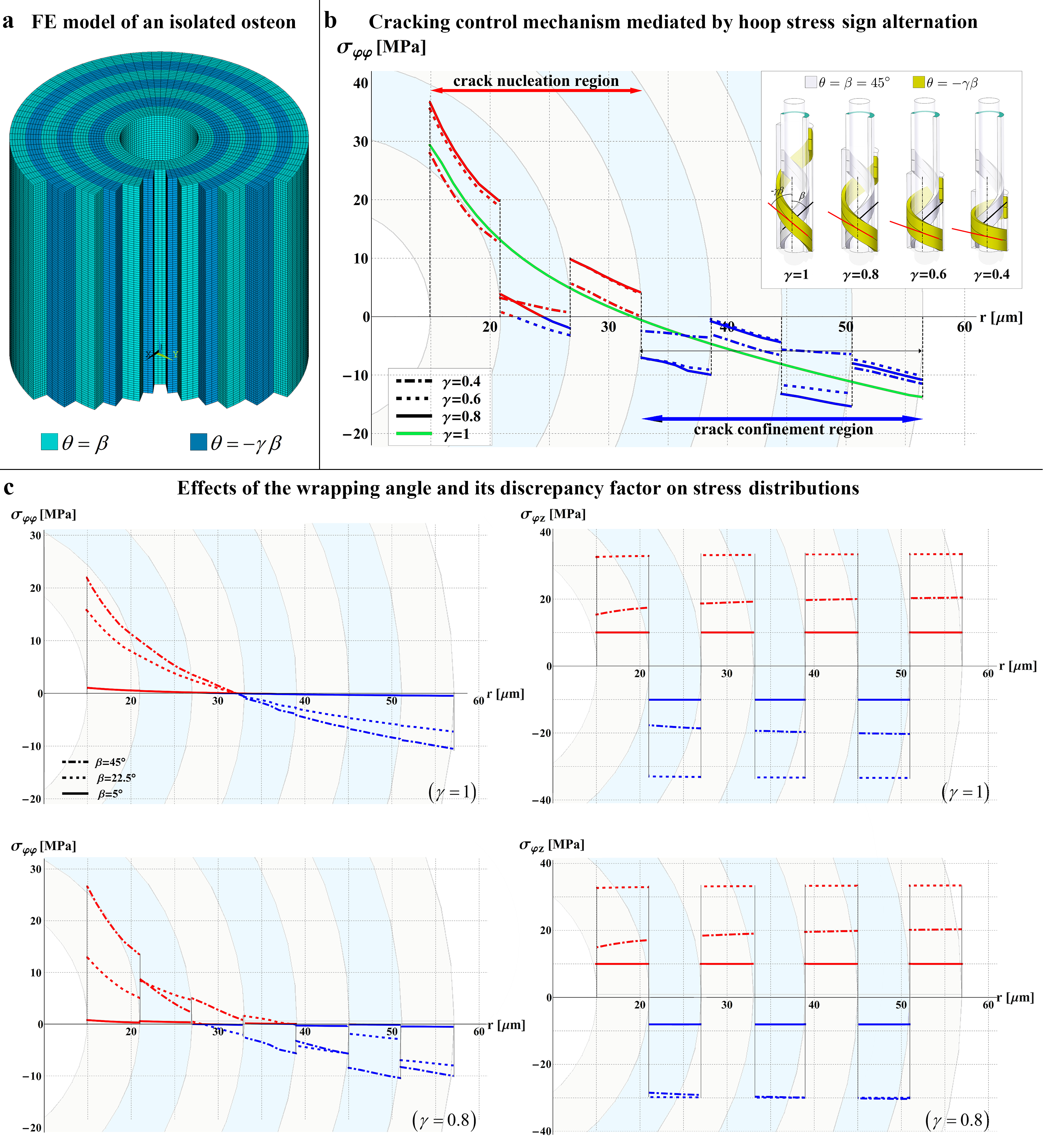}
\caption{\textbf{a)} FE monoclinic model of a secondary osteon implemented in the ANSYS\textsuperscript{\textregistered} code \cite{Ansys}, comprising seven lamellae that contain counter-wrapped fibres with a possible discrepancy factor $\gamma$ to deviate from the perfectly alternate orientation $\theta=\pm \beta$. \textbf{b)} Hoop stresses developing across osteon lamellae in the case of a prescribed axial contraction $\epsilon_0=-10^{-2}$, locked twisting at the bases and traction-free lateral surfaces, for a fixed wrapping angle $\beta=45^{\circ}$ and different discrepancy factors $\gamma$, e.g. $\gamma=0.4, 0.6, 0.8, 1.0$, which highlight how circumferentially tensed regions, where crack opening is expected, are counterbalanced by the sealing effect of compressed ones, an effect amplified if symmetry is lost. \textbf{c)} Hoop $\sigma_{\varphi \varphi}$ (left) and shear $\sigma_{\varphi z}$ (right) stresses for a wide variety of realistic secondary osteon windings (e.g. $\beta=5^{\circ},22.5^{\circ},45^{\circ}°$), both in the case of perfectly counter-wrapped lamellae (i.e. $\gamma=1$, upper row) and experimentally observed asymmetry (i.e. $\gamma=0.8$, lower row), showing that marked osteon chirality (for instance $\beta=45^{\circ}°$) increases stress peaks, favouring the stress-based crack stopping mechanism and magnifying the required osteocyte shear stimulation. Red and blue colors are used for tensile and compressive stress values, respectively. All results have been obtained using the numerical values reported in Table \ref{table}.}
\label{fig.FEresults}
\end{figure}

To reconstruct a more faithful model of an osteon and verify how the stress state behaves in presence of many adjacent lamellae, a detailed FE model is set up. To obtain results of interest for realistic cases, we hence set the anisotropy ratio $\alpha\approx1.7$ and, to consider a wide spectrum of possible secondary osteon structures, the analyses were performed as both the wrapping angles $\beta$ and the discrepancy factor $\gamma$ varied within intervals of values consistent with those observed experimentally. In particular, the osteon FE model conists of seven concentric lamellae all having a thickness of $6 \mu m$ and  variously wound around a central Haversian canal of radius $15 \mu m$ \cite{Wagermaier2006a, Ascenzi1999, Gupta2006a}, as illustrated in Fig. \ref{fig.FEresults}\hyperref[fig.FEresults]{a}. As in the previous simplified 2-phase model, the transverse isotropy in the helical reference system of the intra-lamellar collagen fibrils has been taken into account averaging over each lamella the fibril orientation and by homogenizing the material properties in the helical system, obtaining monoclinic elasticity in cylindrical coordinates. Furthermore, coherently with experimental evidence highlighting imperfectly symmetrical counter-wrapping microstructures, lamellae have been characterized by angles $\vartheta$ that alternately assume values $\beta$ and $-\gamma \beta$ in contiguous phases. The elastic response of this system was then studied in the static regime, prescribing the same boundary conditions applied to the 2-phase model to replicate the above described third stage that determined the stress conditions leading to intra-osteon damaging. 
\\
The results of numerical simulations are summarized in Fig. \ref{fig.FEresults}. Generalizing the 2-phase model with $\beta=45^{\circ}°$, Fig. \ref{fig.FEresults}\hyperref[fig.FEresults]{b} shows how alternate signs of hoop stresses still develop along the osteon radius and lead to potentially anticipated crack nucleation/propagation regions, dominated by tensile stresses, counterbalanced by crack confinement regions, dominated by compressive circumferential stresses. Results also highlight that the hoop stress profile exhibits marked jumps at the interfaces passing from the perfectly counter-wrapped micro-structural arrangement of the lamellae (i.e. $\gamma=1$) to more realistic situations in which counter-wrapping symmetry is instead lost (i.e. $\gamma<1$). This geometrical "imperfection" further improves the toughening mechanism because, irrespective of the applied load sign (e.g. overall contraction or elongation experienced by the osteon), hoop tensile stresses occurring in one or more selected lamellae open a crack: the latter would thus tend to propagate radially at the crack tip, remaining instead trapped in a confined region by sharp jumps of compressive circumferential stresses that simultaneously occur nearby in adjacent lamellae, creating a sealing effect that closes the crack, slowing down or even halting it by at most deflecting its path circumferentially and thus avoiding catastrophic fracturing \cite{Martin2015}.\\
To broaden the campaign of numerical analyses to different osteon  micro-structures, a wider range of possible reference angles $\beta$ was finally investigated, as shown in Fig. \ref{fig.FEresults}\hyperref[fig.FEresults]{c}. More specifically, hoop stresses are plotted along the osteon radius for both perfect alternate wrapping ($\gamma=1$) and an asymmetrical case ($\gamma=0.8$), highlighting a significant amplification of the compressive and tensile peak values as the reference angle $\beta$ --associated with different secondary osteon types-- increases (e.g. $\beta= 5\degree, 22.5\degree, 45\degree$). This outcome is in line with experimental evidence showing that osteons whose average fibre orientations are around $\pm 45\degree$, for which the hypothesised crack-stopping mechanism turns out to be strengthened, are actually those mainly responsible for bone remodelling, since they are significantly more widely present in compact bone in comparison with the so-called \textit{bright} and \textit{dark} osteons, which exhibit sub-horizontal and sub-vertical fibril dispositions, respectively \cite{Giraud-Guille1988}.\\
Further important considerations can be made regarding the shear stresses $\sigma_{\varphi z}$, which exhibit jumps and changes in sign at the interfaces between adjacent lamellae for both perfectly symmetrical ($\gamma=1$) and asymmetrical ($\gamma=0.8$) configurations of the micro-architecture. The increase of wrapping angle $\beta$ and deviation from symmetry associated to $\gamma<1$, however, produce an amplification of anti-plane shear stress magnitudes. This is crucial for bone mechanobiology, since the inter-lamellar jumps imply a magnification of the net shear stresses experienced by the osteocytes inhabiting the lacunae located exactly at the lamellae interfaces. This type of mechanical stimulus, together with the fluid flow, has been ascertained to play a key role in the signal pathway at the basis of bone remodelling \cite{Verbruggen2014,weinbaum1994,CowinPNAS,you2001,cowin1998amp}.

\subsection{Multiple micro-crack opening, stress-induced barriers to crack propagation and magnification of mechanical stimuli from osteon hierarchy}
\label{numeric2}
\begin{figure}[htbp]
\centering
\includegraphics[width=1\textwidth]{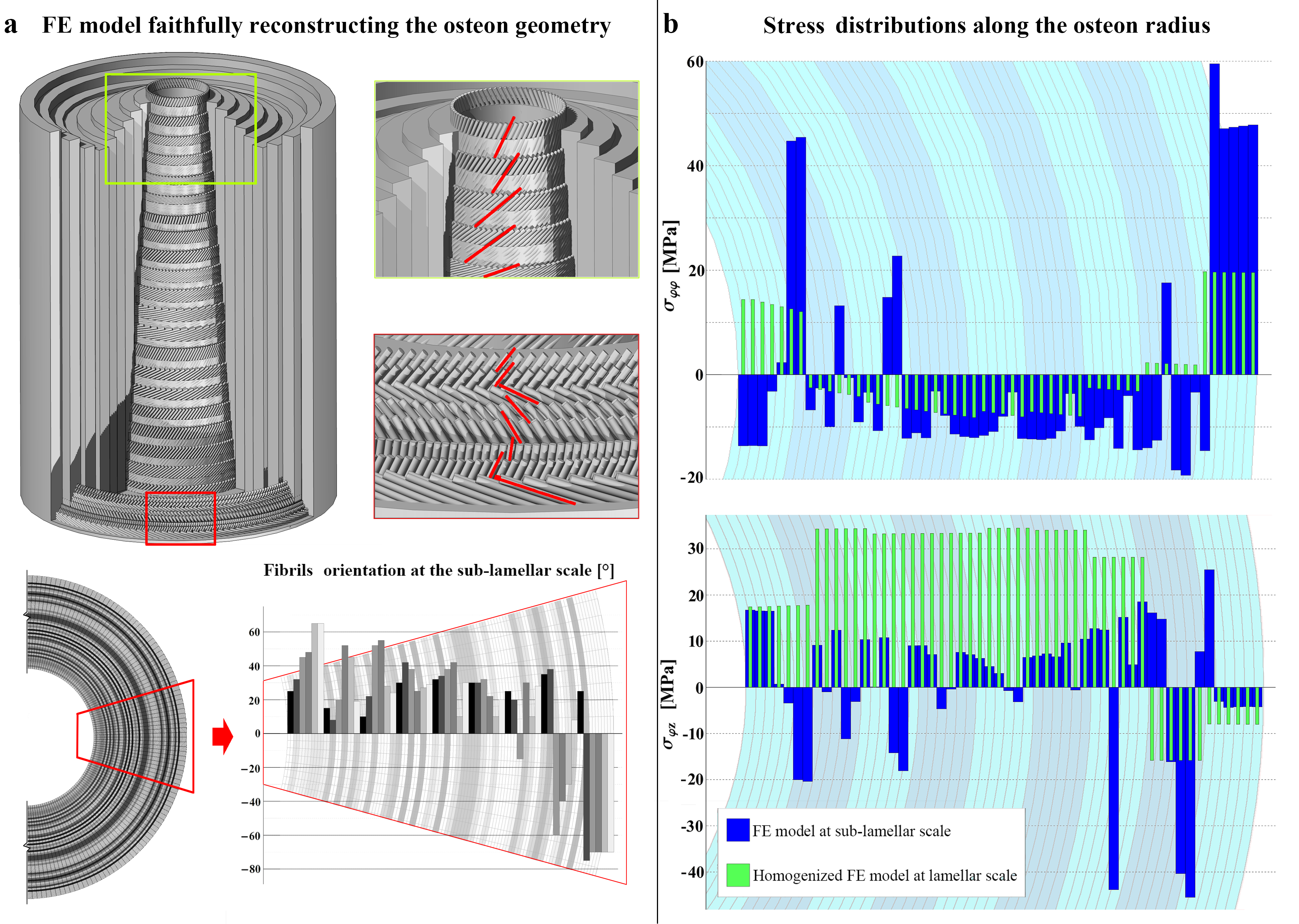}
\caption{\textbf{a)} Geometry of a detailed FE model of a secondary bone osteon, accounting for its sub-lamellar hierarchical organization through the actual arrangement of collagen fibrils reconstructed from experimental data provided by Wagermaier et al. \cite{Wagermaier2006a}. \textbf{b)} Comparison of the results obtained in terms of hoop stresses $\sigma_{\varphi\varphi}$ and shear stresses $\sigma_{\varphi z}$ along the osteon radius, highlighting the growing number of alternate signs within each single lamella and the overall increase of stress peaks for the case in which hierarchy (i.e. sub-lamellar organization) is taken into consideration (in blue), in comparison with the same stress trends when the fibril orientation is averaged over the lamellae thickness (in green).}
\label{fig.FEM_lam_sublam_comparison}
\end{figure}

Raman spectroscopy has widely been used for quantitative analyses of osteons to reveal their micro-structural organization, characterized by packaged bundles of collagen fibres that are helically wrapped around the osteon axis in each lamella, where they form different angles of the same sign (i.e. all clockwise or all counter-clockwise) whose average value is here identified with the already introduced angle $\beta$. To investigate possible effects of osteon hierarchy (i.e. sub-lamellar architecture) on the stress-based crack arresting mechanism, we further enriched the previously described FE osteon model by taking into account the actual arrangement of the fibrils within each lamella. In particular, as shown in Fig. \ref{fig.FEM_lam_sublam_comparison}\hyperref[fig.FEM_lam_sublam_comparison]{a}, the resulting finer numerical model comprises a series of concentric thin (sub-lamellar) cylindrical phases, about $1\mu m$ thick \cite{Ascenzi1999}, made of fibres oriented in a way to exactly reproduce the angle sequence experimentally observed and reconstructed, through an accurate scanning method based on X-ray diffraction, by Wagermaier et al. \cite{Wagermaier2006a} starting from osteon samples. The mechanical response of this more realistic osteon model was then evaluated by applying the same boundary conditions prescribed to the previous cases, finally comparing the results in terms of hoop and shear stresses of the sub-lamellar model with those provided by its simpler counterpart, which was obtained by averaging the fibril orientation over the thickness of each lamella. Results collected in Fig. \ref{fig.FEM_lam_sublam_comparison}\hyperref[fig.FEM_lam_sublam_comparison]{b} show that hierarchy is also explicitly called into play in the stress-based toughening mechanism.
\\
Because of the sub-lamellar organization, both the stress components $\sigma_{\varphi\varphi}$ and $\sigma_{\varphi z}$, which are  responsible for the novel discussed simultaneous crack-opening/crack-halting mechanism and for osteocyte stimulation, respectively, assume marked sign variations even within each single lamella, accompanied by further significant magnification of the stresses. The stress peaks in some regions reach values three or more times greater than the corresponding ones found if the fibrils angles are instead averaged over the lamellae thickness, as highlighted in Fig. \ref{fig.FEM_lam_sublam_comparison}\hyperref[fig.FEM_lam_sublam_comparison]{b}. The sub-lamellar sign alternation and the stress amplification phenomena have a direct impact in increasing the effectiveness of stress-based bone toughening and cell stimulation mechanisms: a first impact is on the optimization of the fracture sealing. Results show how hierarchy leads to cracks nucleating in tensed sub-lamellar regions that will remain more confined in between compressed areas, within the same lamella space. This improves the crack arresting mechanism through the opening of multiple but smaller micro-cracks, locked by circumferential stresses that form real compressive barriers located typically near the Haversian canal and close to the cement line. These give rise to a further stress shielding mechanism limiting the cracks coming from outside the osteon and those directed towards the central blood vessels and nerve fibres, ensuring a controlled diffuse damage that promotes remodelling, while simultaneously improving bone toughness without affecting its stiffness \cite{Ritchie2011,Launey2010}. Additionally, the growth of larger stress peaks (with respect to the non hierarchical, homogenized coarser model) suggests that the sub-lamellar micro-structure works as a sort of gear multiplier of the external loads, thus enhancing the bone sensitivity to the mechanical stimuli and acting in a complementary way to the "strain-amplification" phenomenon envisaged by Han et al. \cite{CowinPNAS,cowin1998amp}. The shear stresses are also amplified by the osteon hierarchical sub-lamellar organization, as seen in Fig. \ref{fig.FEM_lam_sublam_comparison}\hyperref[fig.FEM_lam_sublam_comparison]{b-bottom}: this enhances the mechanical signaling at the lacunae level and increases the probability that osteocytes can be directly affected by the mechanical stimuli, which are crucial in mechano-biological cellular processes to trigger bone remodelling.
  
\section{Beyond the onset of cracks}

\subsection{Numerical simulations of crack progression in osteons}
\begin{figure}[htbp]
\centering	\includegraphics[width=1\textwidth]{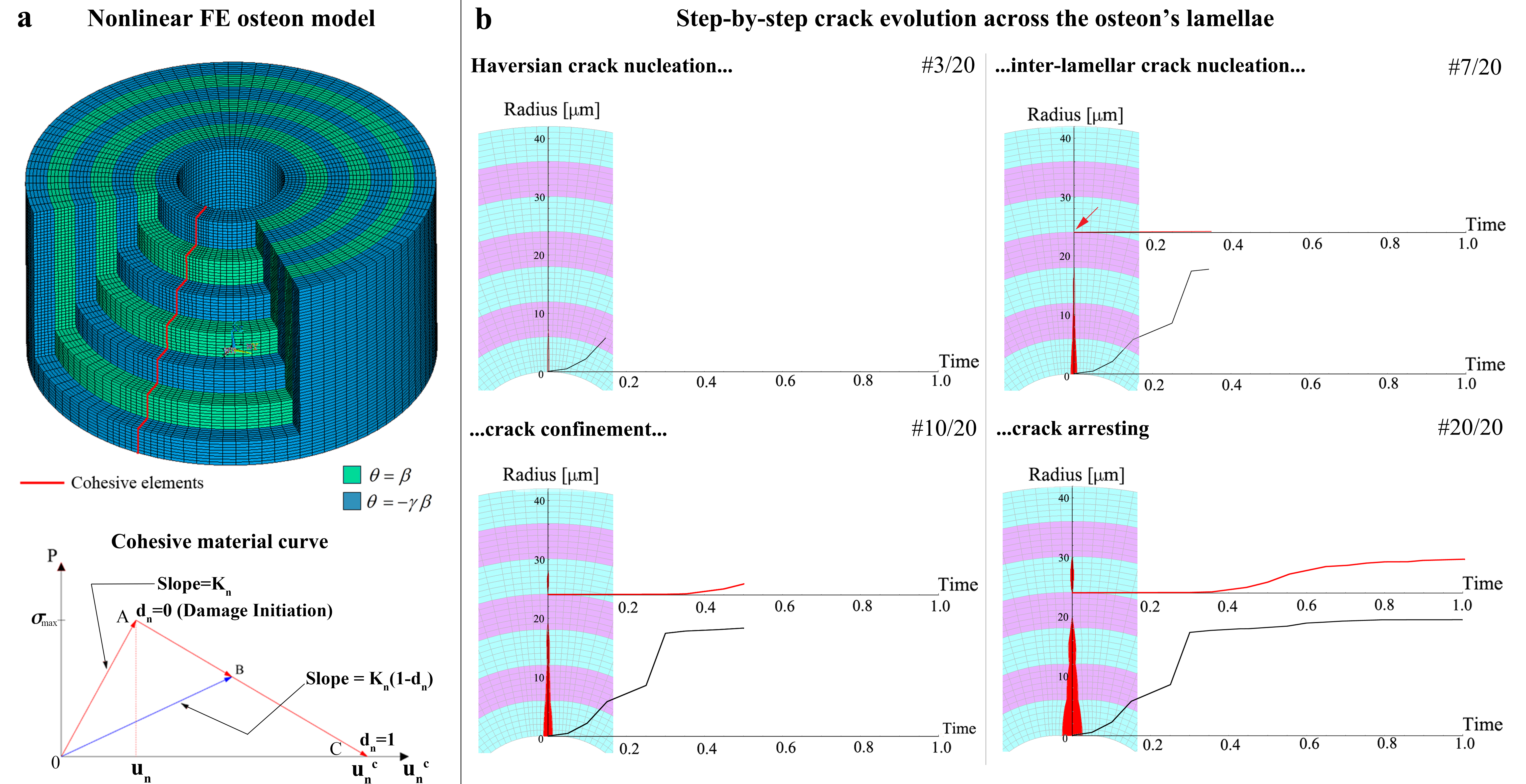}
\caption {\textbf{a)} Nonlinear FE model implemented, at the lamellar scale, to follow the evolution of the stress-induced micro-cracking process occurring within the osteon. The presence of cohesive elements, whose constitutive law is schematically reported in the bottom-left corner, is highlighted in red. \textbf{b)} Progression of the cracking process along the osteon radius, captured at several time-steps in simulations performed beyond the elastic stress analysis. The results were obtained by assuming a thickness equal to $6 \mu m$ for each lamella, innermost radius of $15 \mu m$, osteon height of $100 \mu m$, wrapping angle $\beta=45\degree$ and angle discrepancy factor $\gamma=0.8$, and values reported in Table \ref{table} for the other parameters.}
\label{fig.Cracksequence}
\end{figure}

The analytical solutions and \textit{in silico} FE models presented up to now have investigate how chirality, deviation from symmetry and hierarchy of the osteon micro-structure cooperate to kindle hoop stresses that alternate in sign, leading to the opening of non-propagating micro-cracks, as well as amplified shear stresses stimulating osteocytes located in lacunae at the lamellae interfaces. These results are obtained assuming that the single osteon behaves as a composite cylinder made of elastic monoclinic phases. Although this hypothesis is consistent with the overall bone response when subjected to physiological static or cyclic loads and up to the onset of damage, stresses from elastic analysis can only suggest the locations in which cracks would nucleate and where, at that loading stage, compressive hoop stresses oppose the tensile ones to inhibit their spatial diffusion, in this manner limiting, in case of crack opening, its propagation. However, when a crack opens, the stress state is modified nearby, and eventually interferes with simultaneous stress fields related to other crack nucleations occurring in the vicinity: to determine how cracks evolve and toughness develops requires the performance of simulations in the post-elastic regime, by employing fracture mechanics \cite{RaeisiNajafi2007,Abdel-wahab2012, Ural2013,Giner2014,cornetti2006,taylor2005fracture,pugno2006}. To go beyond elastic analysis and obtain a confirmations of the stress-based toughening mechanism, we studied cracks by employing a simple consolidated cohesive modelling approach. In particular, non-linear FE models were set up by reproducing the osteon micro-structural geometry at both lamellar and sub-lamellar level, using the same elastic properties as above but enriching the elements with the possibility of cracking and updating the model as stress thresholds are reached at any point. Fracture mechanics-based cohesive elements were hence adopted \cite{Ural2013, Giner2014}, which allowed to capture the fracture process by essentially modelling the physical events occurring in the vicinity of the crack through a bi-linear relationship between traction and crack-opening displacement, as shown in Fig. \ref{fig.Cracksequence}\hyperref[fig.Cracksequence]{a}. These cohesive elements were located both on longitudinal planes across the osteon radius and at the lamella-lamella interfaces, where cracks are mainly expected to open and induce debonding or fracture as prescribed tensile strengths are reached. Due to this placement, only radial cracks and possible deviations along the circumferential direction at the lamellae interfaces were allowed during the crack progression simulations, consistently with what is observed in real osteons. 
\\
Overall, the results obtained from fracture analysis, assuming boundary conditions consistent with those prescribed above, confirm the sealing effect, already hypothesised considering the sole elastic regime, kindled by the alternation of the hoop stress sign in the radial direction. In particular, simulations show that, due to the specific orientation of the fibrils, the micro-cracks could both nucleate simultaneously in multiple lamellae and arise in a single region, in the latter case typically starting at points surrounding the Haversian canal. Indeed, close to this region, elastic analysis under the same boundary conditions highlighted the presence of hoop tensile stress peaks. Interestingly, even when a single crack initiates from the innermost regions of the osteon, its propagation tends to develop across few lamellae, being subsequently halted as the load further increases, making room for the nucleation of new cracks in remote zones. These results hence suggest that, even though the specific stress distribution with alternate compressed and tensed regions over the whole osteon radius is obviously perturbed during the transition from the purely elastic regime to the elastic-fracturing one, the stress state tends to dynamically reorganize to reproduce its form that alternates in sign on the residual intact regions. Thus, the spatial variation of stresses realizes the same simultaneous mechanism of crack-opening/stopping for nucleating and confining new cracks far from the early fracture process zone, enormously enhancing the toughness of the bone.\\
An example of this mechanism is summarized in Fig. \ref{fig.Cracksequence}\hyperref[fig.Cracksequence]{b}, where both the crack nucleation and its opening length progression along the osteon radius are illustrated at several time-steps of the simulations performed with reference to a FE model comprising seven lamellae with "imperfect" helicoidal arrangement of the fibrils, which is described by the mean wrapping angle $\theta$ that assumes values $\beta$ and $-\gamma \beta$ in contiguous layers. Due to the stress distribution induced by the growing external load (see Fig. \ref{fig.FEresults}), a first micro-crack arises within the lamella closest to the Haversian canal and then rapidly extends and invades the near lamellar regions. However, after a certain crack size is reached, a second micro-crack nucleates within a lamella, experiencing analogous tensile circumferential stresses, far from the first crack whose propagation is meanwhile stopped. As a result, this produces a slow-down and finally the halt of the first micro-crack and subsequently --following a similar path-- of the second one, too. Both the cracks become wider but no longer grow along the radius as the axial load further increases.
\begin{figure}[htbp]
\centering	
\includegraphics[width=1\textwidth]{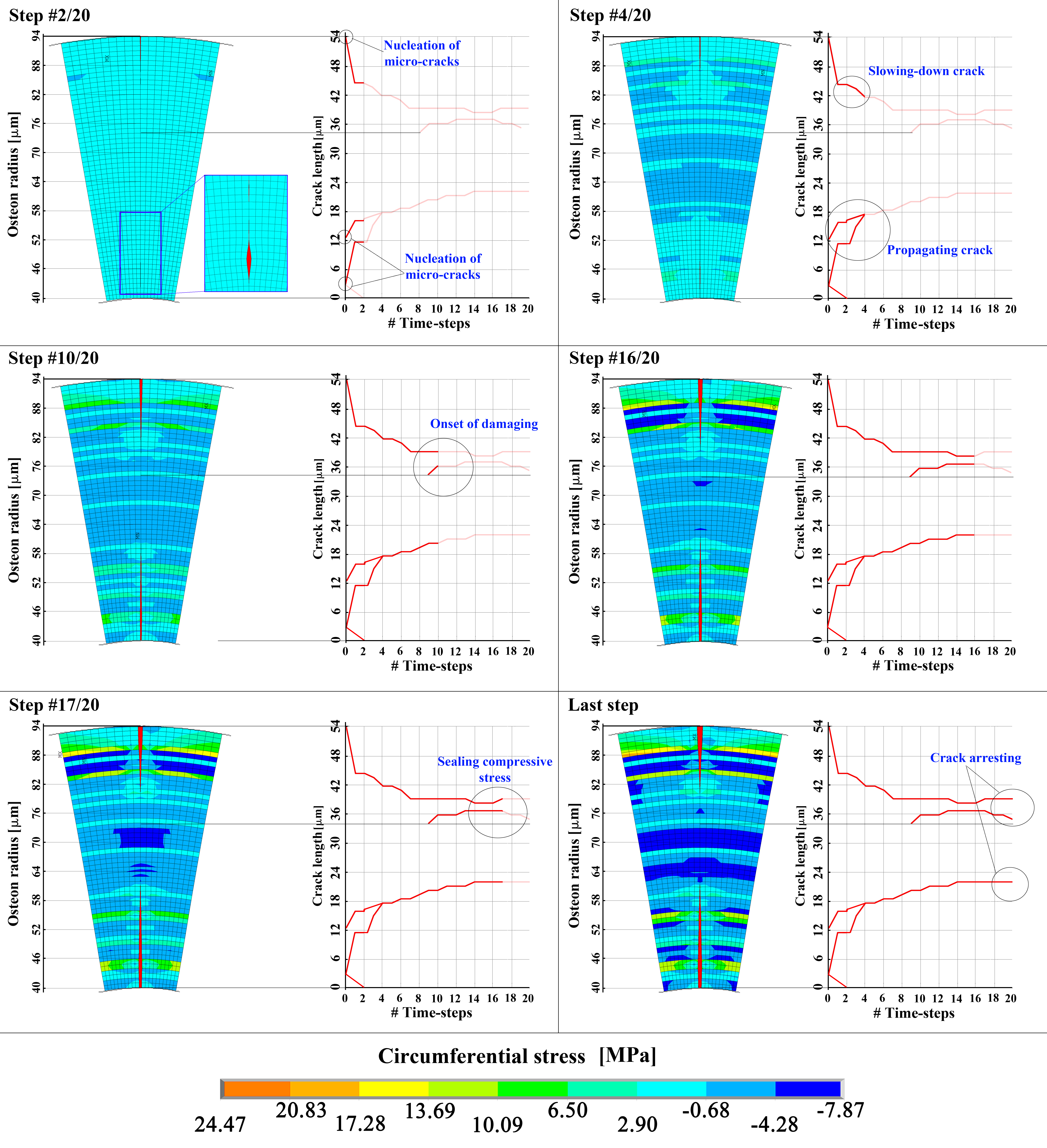}
\caption{Progression of the cracking process along the osteon radius, captured at several time-steps of the simulations performed in the non-linear FE model faithfully reproducing the osteon geometry at the sub-lamellar scale. Consistently with the corresponding elastic analysis, the values of the parameters that have been employed in this case are a thickness equal to $1 \mu m$ for each sub-lamellar layer, an innermost radius of $15 \mu m$, an osteon height of $100 \mu m$ and wrapping angles following the sequence reported in \cite{Wagermaier2006a}. The applied strain is $\epsilon_0=-10^{-2}$. All the other parameters are set according to Table \ref{table}.}
\label{fig.CracksequenceSUBL}
\end{figure}\\
Analogous considerations can be made with reference to the results provided by the step-by-step analysis of the cracking path evolution within a 9-phase osteon described at the sub-labellar scale by following a realistic disposition on the collagen fibrils \cite{Wagermaier2006a}, which are reported in Fig. \ref{fig.CracksequenceSUBL}. In this case, consistently with the elastic stress distributions occurring before any damage occurs (see Fig. \ref{fig.FEM_lam_sublam_comparison}), different cracks arise along the osteon radius, namely at its outermost and innermost lamellae, which are in fact those experiencing the highest levels of tensile stress. For growing external solicitation, in the present case axial contraction, these micro-cracks propagate radially, then slowing down while a further damaging event takes place at an inter-lamellar locus. All the cracks finally halt their progression, remaining confined within small osteon regions and eventually only further thickening laterally.
This kind of behaviour confirms that chirality, deviation from symmetry and hierarchy, characterizing the micro-structure of the osteon, create the alternating stress state in the elastic regime that, as the crack actually nucleates and opens, recurs on the intact parts and works to create mechanical barriers to crack propagation, contributing to confer bone its extraordinary toughness and favouring damage-based remodelling processes. The hypothesised stress-driven mechanism \textit{de facto} explains how micro-crack nucleation, propagation and stopping mechanisms can coexist and collaborate to on the one hand promote necessary mechanobiological activities, and on the other hand, to significantly improve the toughness and the fatigue life of bones under static or cyclic loads \cite{Launey2010,Martin2015,Ritchie2011,Caler1989,Carter1977,Griffin1997,Pattin1996}.

\subsection{Prototyping osteon micro-structures: experimental evidence of the stress-based toughening mechanism}
\begin{figure}[htbp]
\centering	\includegraphics[width=1\textwidth]{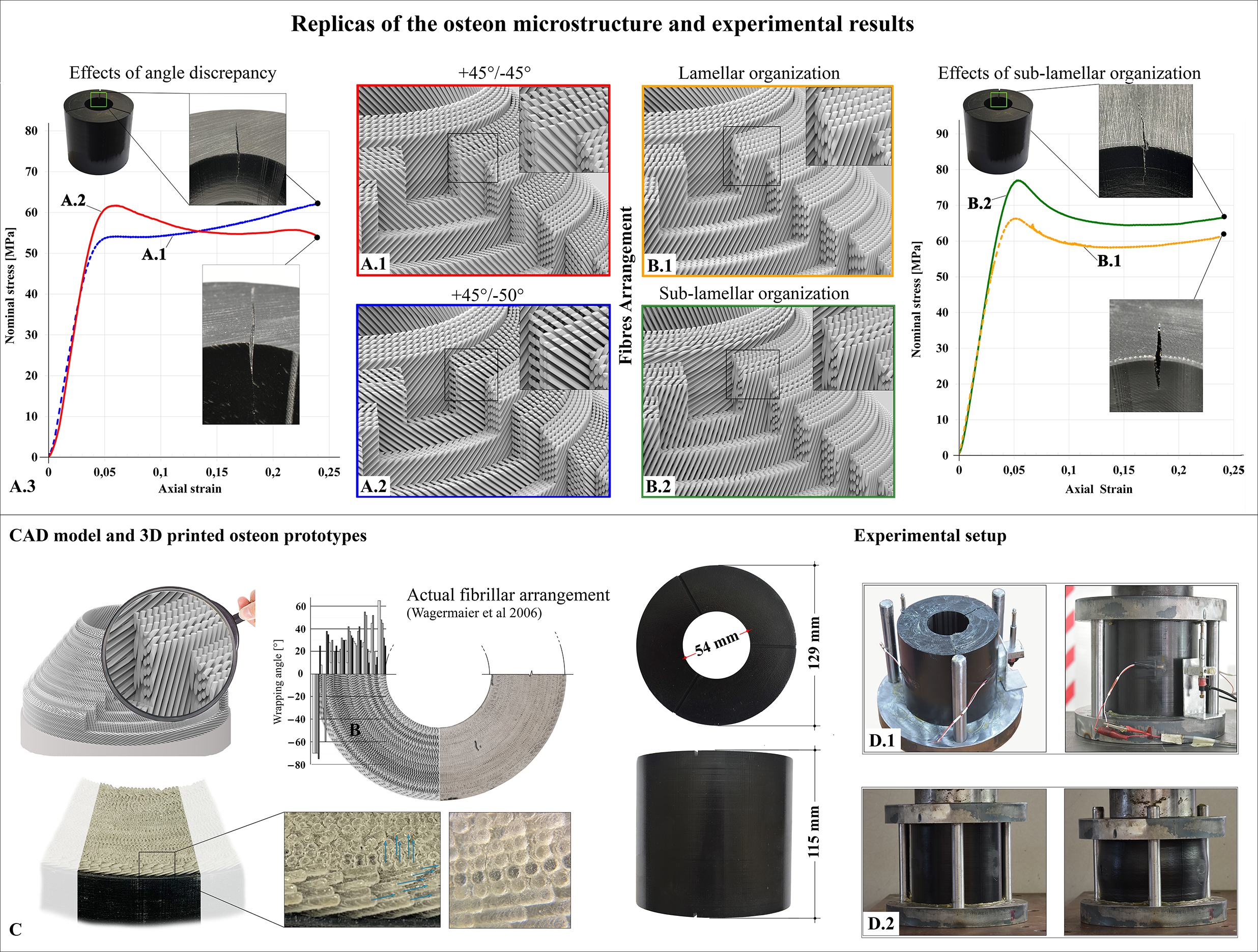}
\caption{3D-printed osteon prototypes and experimental results.
\textbf{A.1-B.2} Details of the fibre arrangement in each lamella for the four osteon samples: prototype with perfectly symmetric counter-wrapped fibres in lamellae with $\beta=\pm45^{\circ}$ (A.1); model with deviation from symmetry, i.e. fibre angles in contiguous lamellae $\beta=+45^{\circ}/-50^{\circ}$ (A.2); osteon prototype with actual fibril orientations averaged over the lamellae thickness (B.1) and replicated faithfully as measured at the sub-lamellar level \cite{Wagermaier2006a}. \textbf{A.3}) Dimensionless axial force versus corresponding engineering longitudinal strain obtained experimentally by testing the first pair of osteon prototypes (a.1 and A.2): note how, starting from a prescribed wrapping angle and a symmetric sequence $\beta=\pm45^{\circ}$ (red curve), the overall toughness of the osteon sample grows as a result of the wrapping angle imperfection (blue curve). The crack in the latter case remains confined to the innermost hollow cylindrical layers (see insets). \textbf{B.3}) Force-strain results comparing the responses of the second pairs of osteon samples (B.1 and B.2): significant toughening (confirmed by the reduced crack extent, illustrated in the inset) is gained if the sub-lamellar hierarchical organization of the osteon is considered (green curve) with respect to the case in which the same micro-structure is instead averaged over the lamellae thickness (yellow curve), consistently with theoretical results. \textbf{C}) Details of both the three-dimensional arrangement of the fibres designed for the CAD model (top-left) and actually realized in the 3D-printed model, following the arrangement in \cite{Wagermaier2006a} (bottom-left and right). \textbf{D1-D2}) Samples instrumented with strain gauges and electromechanical transducers (Linear Variable Differential Transformer, LVDT) to be tested in compression under controlled displacements (D.1) and in undeformed (D.2-left) or deformed (D.2-right) states.}
\label{fig.Exp_img}
\end{figure}

With the aim of providing experimental confirmation of the bone toughening mechanisms described above, brought to light by means of theoretical arguments, we performed some first mechanical tests to indirectly estimate how toughness due to the stress-based crack arresting mechanism varied in detailed replicas of osteons. Due to manufacturing constraints, the model parameters differ from those used to characterize the osteon micro-structure in the theoretical analysis (see Fig. \ref{fig.Exp_img}\hyperref[fig.Exp_img]{C}). More in detail, using high-resolution multi-material additive manufacturing techniques, we \textit{ad hoc} designed and fabricated four polymeric osteon prototypes --hollow cylinders in scale $1000:1$-- consisting of a matrix in which families of helically arranged fibres with different orientations and material properties are embedded, to reproduce both a "regularized" and the "actual" organization of the collagen fibres in osteons. The "regularized" configuration was obtained by grouping fibres in each lamella and disposing all of them following the average wrapping angle (see Fig. \ref{fig.Exp_img}\hyperref[fig.Exp_img]{A.1-A.2}. The "actual" configuration was obtained by reproducing the sub-lamellar micro-structure (see Fig. \ref{fig.Exp_img}\hyperref[fig.Exp_img]{B.1-B.2}. Prototypes were instrumented and tested in laboratory under displacement-controlled uni-axial compression, as shown in Fig. \ref{fig.Exp_img}\hyperref[fig.Exp_img]{D}).
\\
To confirm theoretical results and highlight the role played by the asymmetry of counter-wrapped lamellae and by the hierarchical (sub-lamellar) osteon organization in determining the overall toughness, we fabricated two pairs of osteon prototypes. In the first pair, where the fibre orientations were averaged over each lamella thickness, we realized a perfectly symmetrical model by orienting the fibres in adjacent lamellae symmetrically (i.e. setting $\beta=+45^{\circ}$ and $\beta=-45^{\circ}$), and one with an imperfect symmetry of counter-wrapping lamellae, i.e. $\beta=+45^{\circ}$ and $\beta=-50^{\circ}$. Experimental results shown in Fig. \ref{fig.Exp_img}\hyperref[fig.Exp_img]{A.3} confirmed the analytical and numerical predictions: plotting nominal stress versus axial strain, the toughness modulus indeed increases with the deviation from symmetry (in blue) with respect to the perfectly symmetric lamellar winding (in red). Although the "symmetric model" exhibits a gross yield stress greater than that of the "asymmetrical model", a softening phase follows the stress peak in the symmetric case, while a slight hardening is observed in the case of imperfect symmetry. This result is consistent with the theoretical results, which in the case of $\beta=\pm45^{\circ}$ predict a tensile hoop stress peak at the innermost lamella higher than the one for $\beta=+45^{\circ}/-50^{\circ}$, thus anticipating the crack onset in the asymmetrical case. However, the alternating stresses enhance the sealing effect due to the compressive stresses, in turn slowing down the crack propagation, as shown in Fig. \ref{fig.FEresults}\hyperref[fig.FEresults]{c}. This behaviour is confirmed experimentally, as highlighted by the analysis of specimens cut after the test (see insets in Fig.\ref{fig.FEresults}\hyperref[fig.FEresults]{A.3}), in which cracks that occurred along the radius appeared much more spatially confined in the asymmetrical model than in the symmetrical one.
Furthermore, to investigate the effects of structural hierarchy on toughness, we also fabricated a second pair of osteon prototypes, this time faithfully reproducing the helical counter-wrapped arrangement of the sub-lamellar fibrils of osteons, as reported in \cite{Wagermaier2006a}. In particular, we used as control the sample obtained by homogenizing the actual fibril orientations at the scale of each lamella, the second model instead replicating the actual different sub-lamellar orientations, consistently with the input already implemented to perform the numerical simulations described above. Again, force-strain plots showed how the osteon hierarchy (i.e. the micro-structural arrangements of the fibres at sub-lamellar level) plays a key role in conferring enhanced toughness to bone, as emerges comparing the responses of regularized (yellow curve) and actual (green curve) micro-structures of the osteon samples (see Fig. \ref{fig.Exp_img}\hyperref[fig.Exp_img]{B.3}), with a percentage toughness gain of about $20\%$, here measured by integrating the stress-strain curve not up to the point of failure but at the maximum strain of $25\%$ provided by the testing machine, thus underestimating the actual toughness gain. Accordingly, width and extension of cracks in the two cases (see insets in Fig. \ref{fig.Exp_img}\hyperref[fig.Exp_img]{B.3}) highlighted that the actual sub-lamellar micro-structure slowed down crack propagation more effectively than in the regularized osteon sample, thus indirectly confirming the stress-based toughening mechanism suggested by the theoretical analysis.\\

\section{Conclusion}

At present, bone mechanobiology is only partially understood, with the consequence that many fundamental questions remain still unanswered, including challenges related to ageing-induced bone deterioration and to adult and paediatric diseases. One unsolved enigma in bone is how diffused micro-cracks, necessary to trigger bone growth and remodelling in response to mechanical stimuli at the osteon scale, do not propagate in such a stiff material. It is indeed ascertained that interfaces, sacrificial bonds and material discontinuities alone are insufficient to avoid catastrophic fracture, and thus to provide a definitive response to this open issue.
Starting from exact elastic solutions and accurate numerical simulations, we have demonstrated --with both theoretical arguments and experiments performed on \textit{ad hoc} fabricated 3D-printed prototypes faithfully replicating osteon micro-structures-- that a previously unknown crack-arresting mechanism, mediated by hoop stresses that are alternating in sign, occurs across the lamellae as a result of the cooperation across the scales of the helical counter-wrapped arrangement of the anisotropic fibre bundles, the asymmetry of wrapping angles in contiguous lamellae, and the hierarchical (sub-lamellar) organization of the osteon architecture. We have shown that circumferential stresses with alternate signs balance the opening of new cracks in tensed regions --required for driving nutrient flows, and activating bone tissue remodelling and self-repair processes-- and their arrest in compressed areas, preventing irreversible fractures. Simultaneously, anti-plane shear stresses also appear to be amplified by the sub-lamellar fibre organization, contributing with the interstitial fluid flows to stimulate osteocytes located in the lacunae. We infer that these two complementary and synergistic effects%--which also provide an answer to the provocative question by D. Taylor "Why are your bones not made of steel?" \cite{taylor2010your}-- % 
 create a perfect mechanobiological mechanism, able to operate both in compressive or tensile conditions, hence permitting the functioning in static regimes as well as when bone undergoes cyclic loads. We have also highlighted that the novel stress-driven bone toughening mechanism and the associated phenomenon of mechanical stimulation of osteocytes provided by the presence and amplification of shear stresses, also optimized by deviation from symmetry and hierarchy, are both triggered by the concurrent existence of alternating wrapped structures in the osteon (chirality) and anisotropy (i.e. $\alpha>1$). The change in sign of the hoop stresses and the related crack-opening/-arresting phenomena --at the basis of the bone remodelling-- switch off whether $\alpha=1$ (isotropy of each lamella) or absence of opposite signs of the wrapping angles in adjacent lamellae (e.g. $\beta=0$). It is then worth noticing that osteoporosis, including the effects experienced by astronauts involved in medium/long-term space flights due to microgravity conditions \cite{cowinmicrog,wang2022mechanical}, has been recently demonstrated to be associated to a progressive reduction of the osteon overall anisotropy ratio with respect to healthy bones, combined with a decrease of the level of mineral bone density. This can reach a critical condition of an only partially reversible recovery of the bone's physiological status, for which the behaviour of the osteon tends to finally become isotropic \cite{Ritchie2011,Sabet,cardoso2015changes}. This suggests that the alternating stress mechanism and the obtained results could be exploited to conceive new strategies for personalized regenerative medicine to promote and accelerate the solution to bone tissue diseases and degenerative disorders, as well as to design novel bio-inspired hierarchical, self-healing and ultimately self-remodelling materials.

\section{Methods}
\subsection{FE elastic model of the osteon}
Several parametric procedures have been developed \textit{ad hoc} by using the APDL interface in order to automatically generate the FE models reproducing the geometrical and constitutive properties of the osteon and to analyse their mechanical response with the aid of the ANSYS\textsuperscript{\textregistered} code \cite{Ansys}.\\
In particular, the implemented geometry of the FE osteon model at the lamellar scale consists of a FGMC, containing seven concentric layers of thickness $6 \mu m$, wound around a central Haversian canal having a radius of $15 \mu m$. The height to diameter ratio for the whole hollow cylinder was set to be $\approx 1$, this size being sufficiently representative of the osteon and such that end effects were avoided. The FE model was then obtained by discretizing the solid by means of about 378,000 standard 8-node hexahedral elements (SOLID185) and about 390,000 nodes. The implemented parametric modelling allowed to characterize the constitutive properties in an independent way for each lamella, by setting appropriate elastic constants $C^{cyl}_{ij}$ in the cylindrical reference system $\left\{r,\varphi,z\right\}$ and by making use of local helical coordinates to describe the transverse isotropy of the single fibrils, whose direction was traced by the wrapping angle $\vartheta$ alternately assuming values $\beta$ and $-\gamma \beta$ in contiguous phases. Coherently with the analytical model, we studied the mechanical response of the overall FE system by prescribing vanishing tractions at the lateral surfaces and a longitudinal contraction $\epsilon_{0}=-10^{-2}$ through the imposition of a uniform axial displacement $u_{z}$ at the upper and lower bases, at which the twisting was also locked by imposing a circumferential displacement $u_{\varphi}=0$.\\
By following an analogous approach, we set up the more detailed FE osteon model, described in the main text, containing instead nine lamellae whose micro-structure was taken into account by reproducing the actual fibril orientations experimentally observed and measured in \cite{Wagermaier2006a} for each sub-lamellar layer. In this case, the mesh of the model generated 48600 elements and 54450 nodes.

\subsection{FE simulations of crack evolution within osteons}
With reference to the nonlinear FE simulations performed to follow the whole evolution of the cracks within the osteon, a standard fracture cohesive model was implemented and the parameters were chosen as high as possible to properly satisfy numerical convergence, by considering that the crack formation was influenced predominantly by a single \textit{mode I}. The implemented geometry was meshed with about 244,000 elements and about 292,000 nodes in the case of the lamellar description and with 108680 elements and 103950 nodes in the case of the sub-lamellar (hierarchical), more detailed model.

\subsection{3D prototyping and mechanical testing of osteon-inspired polymeric models}
The four prototypes of the osteon were designed \textit{ad hoc} and manufactured through a high-resolution and multi-material 3D printing machine (Stratasys J750). They consist in polymeric multi-layered hollow cylinders reinforced with different families of helically organized fibres, replicating the main features of the osteon architecture (see Fig. \ref{fig.Exp_img}\hyperref[fig.Exp_img]{A.1-B.2, C}). More in detail, the  multi-material 3D printer Stratasys J750 allowed to fabricate prototypes with locally defined constitutive properties. Thus, by using the resinous material VeroClear\texttrademark 810 to obtain fibers with elastic stiffness around $2-3$ GPa and a combination of VeroBlack\texttrademark 870 and AgBLK30\texttrademark  to obtain a matrix with a Young's modulus of about $1.1-1.7$ GPa, it was possible to fabricate the desired composite structures with an anisotropy ratio $\alpha$ of about $ 1.7$, consistently with the values reported in the literature for bones \cite{yoon2008} and used in this work in both theoretical analysis and FE numerical procedures. In particular, to build up the sample faithfully reproducing the counter-wrapped arrangement of the osteon fibrils at the sub-lamellar level, a very accurate, fine CAD model was implemented, which included concentric lamellae, each made of six sub-lamellae comprising fibers whose number and wrapping angle varied parametrically from the innermost to the outermost layer to ensure the same volumetric fraction of fibres along the radius. This is necessary to quantitatively compare the test results in the various cases analysed. The accuracy of the prototype structures, highlighted in Fig. \ref{fig.Exp_img}, was guaranteed by the high-resolution printing process with a slice thickness of about $30\mu m$ and a spatial resolution of about $2\mu m$.\\
The four multi-layered cylindrical structures were then tested under axial contraction by means of a servo-hydraulic machine ITALSIGMA (having a loading capability of $3000$kN in compression and a maximal cross-head displacement of $75$mm), with the aim to evaluate their elastic response, yield strength and post-elastic behaviour. The latter is related to toughness and damaging phenomena, which were observed at the end of each test by cutting some slices from the tested samples to find any cracks present. In particular, several strain gauges (positioned circumferentially) and LVDT sensors (positioned longitudinally) were used to control the accuracy of both circumferential and axial strains during the uni-axial compression tests.

\paragraph*{Acknowledgments} 
MF, AC, SP and NMP acknowledge the support of the Italian Ministry of Education, University and Research (MIUR) through the grant PRIN-20177TTP3S.  FB and NMP also acknowledges the support of the FET Open “Boheme” grant no. 863179. ARC acknowledges the MIUR support of the grant PON-AIM1849854-1.

\newpage
%\appendix
%\include{AppendixA}
%\include{AppendixB}
\section*{Supplementary Information} \label{SB2}
\renewcommand{\theequation}{S.\arabic{equation}}
  % redefine the command that creates the equation no.
\setcounter{equation}{0}  % reset counter
%%%%%%%%%%%%%%%%%%%%%%%%%%%%%%%%%%%%%%%%%%%

\section*{Analytical modelling of an isolated osteon}
\label{AppendixA}
To model the mechanical behavior of an isolated osteon in response to selected physiological constraints and loading conditions, let us consider a cylindrical reference system $\left\{r,\varphi,z\right\}$ having the origin placed at the basis of the whole FGMC model, say at the center of the Harvesian canal. Then, in the light of the above described characteristic micro-structure, the osteon exhibits a plane of material symmetry $\varphi-z$, which allows to assume a cylindrically monoclinic anisotropy \cite{Fraldi2002,cowin2013b,cowin1987}, here retrieved by considering a transverse isotropy in a local helical reference system fixed with respect to the plane orthogonal to the axis of the fibres and then transforming all the quantities of interest in the global cylindrical coordinate system. 
Then, the linear elastic law for the generic $j$-th phase ($j=1,...,n$) can be written as
\begin{equation}
\label{eq.costi}
\boldsymbol{\sigma}=\textbf{C}:\boldsymbol{\epsilon}
\end{equation}
where the stress $\boldsymbol{\sigma}$, the strain $\boldsymbol{\epsilon}$ and the elasticity matrix $\textbf{C}$, in the Voigt notation, read respectively as
\begin{equation}
\label{eq.sistemel}
\boldsymbol{\sigma}=
\left[
\begin{array}{c}
 \sigma _{rr} \\
 \sigma _{\varphi \varphi } \\
 \sigma _{zz} \\
 \sigma _{\varphi z} \\
 \sigma _{rz} \\
 \sigma _{r\varphi}
\end{array}
\right],
\boldsymbol{\epsilon}=
\left[
\begin{array}{c}
 \epsilon_{rr} \\
 \epsilon_{\varphi\varphi} \\
 \epsilon_{zz} \\
 \epsilon_{\varphi z} \\
 \epsilon_{rz} \\
 \epsilon_{r\varphi}
\end{array}
\right],
\textbf{C}=
\left[
\begin{array}{cccccc}
 C_{11} & C_{12} & C_{13} & C_{14} & 0 & 0 \\
 C_{12} & C_{22} & C_{23} & C_{24} & 0 & 0 \\
 C_{13} & C_{23} & C_{33} & C_{34} & 0 & 0 \\
 C_{14} & C_{24} & C_{34} & C_{44} & 0 & 0 \\
 0 & 0 & 0 & 0 & C_{55} & C_{56} \\
 0 & 0 & 0 & 0 & C_{56} & C_{66}
\end{array}
\right]
\end{equation}
with
\[
\boldsymbol{\epsilon}=
\left[
\begin{array}{c}
 u_{r,r},
 \tfrac{u_{\varphi,\varphi}+u_r}{r},
 u_{z,z},
 \tfrac{u_{\varphi ,z}+r u_{\varphi,z}}{2r},
 \tfrac{u_{r,z}+u_{z,r}}{2},
 \tfrac{ru_{\varphi ,r}+u_{r,\varphi}-u_{\varphi }}{2r}
\end{array}
\right]^T
\]
Then, by considering both the axis-symmetry of the osteon geometry and the prescribed boundary conditions of interest (i.e. traction-free lateral surfaces and imposed uniform axial strain with locked twist at the cylindrical bases), the displacement field can be assumed as independent from the variable $\varphi$, so that the solution for the $j$-th phase can be sought in the form:
\[
u_r=u_r(r,z) \quad u_\varphi=u_\varphi(r,z) \quad u_z=u_z(r,z)
\]
the corresponding shear strain components being:
\[
\gamma_{\varphi z}=2 \epsilon_{\varphi z} \quad \gamma_{r z}=2 \epsilon_{r z} \quad \gamma_{r \varphi}=2 \epsilon_{r  \varphi}
\]
Therefore, the constitutive relation \eqref{eq.costi} can be rewritten as follows:
\begin{equation}
\label{eq.sistel2}
\left[
\begin{array}{c}
 \sigma _{rr} \\
 \sigma _{\varphi \varphi } \\
 \sigma _{zz} \\
 \tau _{\varphi z} \\
 \tau _{rz} \\
 \tau _{r\varphi}
\end{array}
\right]=
\left[
\begin{array}{cccccc}
 C_{11} & C_{12} & C_{13} & C_{14} & 0 & 0 \\
 C_{12} & C_{22} & C_{23} & C_{24} & 0 & 0 \\
 C_{13} & C_{23} & C_{33} & C_{34} & 0 & 0 \\
 C_{14} & C_{24} & C_{34} & C_{44} & 0 & 0 \\
 0 & 0 & 0 & 0 & C_{55} & C_{56} \\
 0 & 0 & 0 & 0 & C_{56} & C_{66}
\end{array}
\right]
\cdot
\left[
\begin{array}{c}
 u_{r,r} \\
 \tfrac{u_r}{r} \\
 u_{z,z} \\
 u_{\varphi ,z} \\
 u_{r,z}+u_{z,r} \\
 u_{\varphi ,r}-\tfrac{u_{\varphi }}{r}
\end{array}
\right]
\end{equation}
With respect to the stress components, physical arguments allow to assume that they can be found as independent from the $z$ variable, this implying that the following set of differential equations holds true:
\begin{equation}
\label{eq.diffsel}
\begin{cases}
u_{r,rz}=0\\
u_{r,z}=0
\end{cases}
\begin{cases}
u_{\varphi,zz}=0\\
u_{\varphi,rz}-\dfrac{u_{\varphi,z}}{r}=0
\end{cases}
\begin{cases}
u_{z,zz}=0\\
u_{r,zz}+u_{z,rz}=0
\end{cases}
\end{equation}
By integrating them, the displacements will take the general form:
\begin{align}
\label{eq.dformel}
&u_{r}=U(r) \notag \\
&u_{\varphi}=V(r)+\phi r z  \\
&u_{z}=W(r)+\epsilon_0 z \notag
\end{align} 
where $U(r)$,$V(r)$ and $W(r)$ are unknown functions and $\phi$ and $\epsilon_0$ are constants to be determined by imposing boundary conditions to the system. By substituting the expressions \eqref{eq.dformel} into the Cauchy equilibrium equations that stresses have to obey, i.e. $div \boldsymbol{\sigma}=\textbf{0}$, the differential system in the unknown functions $U(r)$,$V(r)$,$W(r)$ is finally obtained:
\begin{equation}
\label{eq.sistufel}
\begin{sistema}
C_{11} r^2 U''(r)+C_{11} r U'(r)-C_{22} U(r)+(C_{13}-C_{23}) r \epsilon _0+
r^2 \phi(2 C_{14}-C_{24})=0\\
\\
C_{66} \left[r \left(r V''(r)+V'(r)\right)-V(r)\right]+C_{56} r \left(r W''(r)+2 W'(r)\right)=0\\
\\
C_{56} V''(r)+C_{55} \left(W''(r)+\dfrac{W'(r)}{r}\right)=0
\end{sistema}
\end{equation}
Therefore, by solving the first of \eqref{eq.sistufel} to find $U(r)$ and by then integrating the second and the third ones with respect to the unknown functions $V(r)$ and $W(r)$ and substituting the results in \eqref{eq.dformel}, algebraic manipulations lead to obtain the displacement components for the $j$-th phase as:
\begin{align}
\label{eq.fieldurel}
u_{r}^\textup{(j)}=& U_1^{\textup{(j)}}(r^{\lambda^\textup{(j)} }+r^{-\lambda^\textup{(j)} })+i U_2^{\textup{(j)}}(r^{\lambda^\textup{(j)} }+r^{-\lambda^\textup{(j)} })+h_1^{\textup{(j)}}\epsilon_0^{\textup{(j)}}r+h_2^{\textup{(j)}}\phi^{\textup{(j)}}r^2 \\
\label{eq.fieldufel}
u_{\varphi}^\textup{(j)}=&r V_1^{\textup{(j)}}+\frac{V_3^{\textup{(j)}}}{r}+r z \phi ^{\textup{(j)}}+V_0^{\textup{(j)}} \\
\label{eq.fielduzel}
u_{z}^\textup{(j)}=&\frac{V_0^{\textup{(j)}} \ln (r)}{\omega_1^\textup{(j)} }-\frac{2 V_3^{\textup{(j)}}}{\omega_2^\textup{(j)}  r}+W_0^{\textup{(j)}}+ \epsilon _0^{\textup{(j)}}\,z  
\end{align}
where
\begin{align*}
&h_1^\textup{(j)}=\tfrac{C_{23}^{\textup{(j)}}-C_{13}^{\textup{(j)}}}{C_{11}^{\textup{(j)}}-C_{22}^{\textup{(j)}}} & &h_2^\textup{(j)}=\tfrac{C_{24}^{\textup{(j)}}-2C_{14}^{\textup{(j)}}}{4C_{11}^{\textup{(j)}}-C_{22}^{\textup{(j)}}} & &\lambda^\textup{(j)}=\tfrac{\sqrt{C_{22}^\textup{(j)}}}{\sqrt{C_{11}^\textup{(j)}}} \\
&\omega_1^\textup{(j)}=\tfrac{C_{56}^\textup{(j)}}{C_{66}^\textup{(j)}} & &\omega_2^\textup{(j)}=\tfrac{C_{55}^\textup{(j)}}{C_{56}^\textup{(j)}}
\end{align*}
By substituting these expressions in the system \eqref{eq.sistel2}, it can be seen that $W_0^{\textup{(j)}}$ and $V_1^{\textup{(j)}}$ are constants associated with rigid body motions, so \eqref{eq.fieldufel} and \eqref{eq.fielduzel} simplify as:
\begin{align}
\label{eq.fielduf2el}
u_{\varphi}^\textup{(j)}=&\frac{V_3^{\textup{(j)}}}{r}+r z \phi ^{\textup{(j)}}+V_0^{\textup{(j)}} \\
\label{eq.fielduz2el}
u_{z}^\textup{(j)}=&\frac{V_0^{\textup{(j)}} \ln (r)}{\omega_1^\textup{(j)} }-\frac{2 V_3^{\textup{(j)}}}{\omega_2^\textup{(j)}  r}+ \epsilon _0^{\textup{(j)}}\,z 
\end{align}
Solutions \eqref{eq.fieldurel}, \eqref{eq.fielduf2el}, \eqref{eq.fielduz2el} and the relationship \eqref{eq.sistel2} then allow to derive the stress components and, without loss of generality and with the displacement-prescribed problem in mind, the following conditions at the bases of the osteon can be written:
\begin{equation}
\label{eq.dispcondel}
\begin{cases}
u_z^\textup{(j)}|_{z=L}=D_0 \\
u_\varphi^\textup{(j)}|_{z=L}=\phi_0 r
\end{cases}
\forall j \in \left\{0,1,...n\right\}
\end{equation}
where $L$ is the total length of the hollow cylinder and $D_0$ and $\phi_0$ are respectively a prescribed displacement value and a prescribed twisting angle, shared by all the phases. As a consequence, by solving \eqref{eq.dispcondel}, an uniform deformation in $z$ direction $\epsilon_0$ and a unitary angle of rotation $\phi$ for all the phases are obtained as functions of $D_0$ and $\phi_0$, that is:
\begin{equation}
\begin{aligned}
\label{eq.epsfiel}
\epsilon_0^\textup{(j)}=\tfrac{D_0}{L} \\
\phi^\textup{(j)}=\tfrac{\phi_0}{L}
\end{aligned}
\qquad \forall j \in \left\{1,...n\right\}
\end{equation}

In order to take into account the simplest form in which the elastic moduli of the monoclinic stiffness matrix $\textbf{C}$ can be expressed, we recall that the structure of the osteon is such that each lamella exhibits transverse isotropy with respect to its own helical coordinate system, the plane of isotropy being orthogonal to the local fibres axis and the passage between the overall cylindrical reference system and the local helical one being mediated by the transformation process illustrated in Appendix \ref{AppendixA}. In detail, the five independent elastic moduli for a generic $j$-th phase can be referred to the helical coordinate system $\left\{r,t,c\right\}$, characterized by the unit vectors $\left\{\textbf{e}_r,\textbf{e}_t,\textbf{e}_c\right\}$: in this system, the elastic compliance tensor $\textbf{S}^{hel}$ reads as:
\begin{equation}
\textbf{S}^{hel}=\left[
\begin{array}{cccccc}
 \tfrac{1}{E_{rr}} & -\tfrac{\nu_{rc}}{E_{cc}} & -\tfrac{\nu_{rt}}{E_{tt}} & 0 & 0 & 0 \\
 -\tfrac{\nu_{rc}}{E_{cc}} & \tfrac{1}{E_{cc}} & \tfrac{\nu_{tc}}{E_{tt}} & 0 & 0 & 0 \\
 -\tfrac{\nu_{rt}}{E_{tt}} & -\tfrac{\nu_{tc}}{E_{tt}} & \tfrac{1}{E_{tt}} & 0 & 0 & 0 \\
 0 & 0 & 0 & \tfrac{1}{G_{rc}} & 0 & 0 \\
 0 & 0 & 0 & 0 & \tfrac{1}{G_{rt}} & 0 \\
 0 & 0 & 0 & 0 & 0 & \tfrac{1}{G_{tc}}
\end{array}
\right]
\label{eq.complian}
\end{equation}
where we set
\begin{equation}
E_{rr}=E_{cc}=E, \quad E_{tt}=E_t=\alpha E, \quad \nu_{rc}=\nu, \quad \nu_{rt}=\nu_{tc}=\nu_t, \quad G_{rc}=G=\frac{E}{2(1+\nu)}, \quad G_{rt}=G_{tc}=G_t=\eta G
\label{eq.relelcost}
\end{equation}

denoting with $E$ and $\nu$ the Young's modulus and Poisson's coefficient in the plane of isotropy, $\alpha \geq 1$ being a dimensionless parameter describing the anisotropy ratio (the ratio between the elastic modulus along the fibres direction and the Young modulus in the plane of isotropy whose normal is coaxial with the fibres), $\nu_t$ representing the Poisson's ratio in the anisotropy planes and the coefficient $\eta>0$ relating the shear moduli, positive definitiveness of the elastic energy implying the following restrictions:
\begin{equation}
-1<\nu<1, \qquad -\frac{\sqrt{\alpha}(1+\nu)}{\sqrt{2}}<\nu_t<\frac{\sqrt{\alpha}(1+\nu)}{\sqrt{2}} \quad \cap \quad -\sqrt{\alpha}<\nu_t<\sqrt{\alpha}
\label{eq.nirel}
\end{equation}
From the relations above, the following elastic stiffness coefficients can be finally derived in the helical coordinate system:
\begin{align}
\label{eq.consthel}
&C^{hel}_{11}=C^{hel}_{22}=\frac{E(\nu^2-\alpha)}{(1+\nu)[\alpha(\nu-1)+2\nu_t^2]} \notag \\
&C^{hel}_{12}=-\frac{E(\nu_t^2-\alpha \nu)}{(1+\nu)[\alpha(\nu-1)+2\nu_t^2]} \notag \\
&C^{hel}_{13}=C^{hel}_{23}=\frac{E\alpha \nu}{\alpha-\alpha \nu -2\nu_t^2} \\
&C^{hel}_{33}=\frac{E\alpha^2(\nu-1)}{\alpha(\nu-1)-2\nu_t^2} \notag \\
&C^{hel}_{44}=C^{hel}_{55}=\frac{E \eta}{2(1+\nu)} \notag \\
&C^{hel}_{66}=\frac{E}{2(1+\nu)} \notag
\end{align}
Then, by using the relations \eqref{eq.heltocyl} in Appendix \ref{AppendixA} the expression of the stiffness components $C_{ij}$ in the cylindrical coordinate system (from \eqref{eq.consthel}) are finally obtained as a function of the angle $\vartheta$, which represents the orientation of the fibres in a single lamella evaluated with respect to the osteon axis.

\subsection*{Exact solutions for the n-phase osteon model}
By considering the generic case in which there are $n$ arbitrary hollow cylindrical phases and under the above mentioned prescribed boundary conditions, the unknown parameters result:
\begin{equation*}
V_0^\textup{(j)} \quad V_3^\textup{(j)} \quad U_1^\textup{(j)} \quad U_2^\textup{(j)} \quad \phi^\textup{(j)} \quad \epsilon^\textup{(j)}_0 \qquad \forall j \in \left\{1,...n\right\}
\end{equation*}
Hence the total number of unknowns is $6n$ that must equate the total number of boundary conditions of the assigned problem. At the interfaces, the $6(n-1)$ equilibrium and compatibility equations read as:
\begin{equation}
\label{eq.condeln}
\begin{aligned}
\sigma_{rr}^\textup{(j)}|_{r=R_\textup{(j)}}=\sigma_{rr}^\textup{(j+1)}|_{r=R_\textup{(j)}} \\
\sigma_{r \varphi}^\textup{(j)}|_{r=R_\textup{(j)}}=\sigma_{r \varphi}^\textup{(j+1)}|_{r=R_\textup{(j)}} \\
\sigma_{rz}^\textup{(j)}|_{r=R_\textup{(j)}}=\sigma_{rz}^\textup{(j+1)}|_{r=R_\textup{(j)}} \\
u_r^\textup{(j)}|_{r=R_\textup{(j)}}=u_r^\textup{(j+1)}|_{r=R_\textup{(j)}} \\
u_\varphi^\textup{(j)}|_{r=R_\textup{(j)}}=u_\varphi^\textup{(j+1)}|_{r=R_\textup{(j)}}\\
u_z^\textup{(j)}|_{r=R_\textup{(j)}}=u_z^\textup{(j+1)}|_{r=R_\textup{(j)}}
\end{aligned}
\qquad \forall j \in \left\{1,...n-1\right\}
\end{equation}

it can be shown from the second, third, fifth and last of \eqref{eq.condeln} that $\phi^\textup{(j)}=\phi$, $\epsilon^\textup{(j)}=\epsilon_0$ and $V_0^\textup{(j)}=V_3^\textup{(j)}=0$. In general linear elastostatic problems with load-prescribed boundary conditions, the variables $\epsilon_0$ and $\phi$ are treated as additional unknowns, so that the number of constants to be determined becomes $2n+2$ and the previous conditions reduce to:
\begin{equation}
\label{eq.condeln2}
\begin{aligned}
\sigma_{rr}^\textup{(j)}|_{r=R_\textup{(j)}}=\sigma_{rr}^\textup{(j+1)}|_{r=R_\textup{(j)}} \\
u_r^\textup{(j)}|_{r=R_\textup{(j)}}=u_r^\textup{(j+1)}|_{r=R_\textup{(j)}}
\end{aligned}
\qquad \forall j \in \left\{1,...n-1\right\}
\end{equation}
Then, additional conditions require the equilibrium on the internal and external cylindrical boundary surfaces, i.e.
\begin{align}
\label{eq.equilcond}
&\sigma_{rr}^{(1)}|_{r=R_0}=0 \\
&\sigma_{rr}^\textup{(n)}|_{r=R_n}=0 \notag
\end{align}
while, at one of the bases, the equilibrium equation along and about the $z$-direction must be considered. Therefore, without loss of generality, for $z=0$, one has
\begin{align}
\label{eq.zcond}
&\sum^n_{i=1}\int^{2\pi}_0 \int^{R_i}_{R_{i-1}} \sigma^\textup{(i)}_{zz}|_{z=0}\quad rdrd\varphi=F_z \\
&\sum^n_{i=1}\int^{2\pi}_0 \int^{R_i}_{R_{i-1}} \sigma^\textup{(i)}_{\varphi z}|_{z=0}\quad r^2drd\varphi=\mathcal{M}
\end{align}
where $F_z$ and $\mathcal{M}$ are the total axial force and torque moment applied at $z=0$. In this way, the system is closed and can be solved. Recasting the \eqref{eq.condeln2} in the form
\begin{align}
\label{eq.condeln3}
&U^\textup{(j)}_1 A^\textup{(j)}_{5L}+U^\textup{(j+1)}_1 A^\textup{(j)}_{6L}+U^\textup{(j)}_2 A^\textup{(j)}_{7L}+U^\textup{(j+1)}_2 A^\textup{(j)}_{8L} + \epsilon_0 A^\textup{(j)}_{9L} + \phi A^\textup{(j)}_{10L} + E^\textup{(j)}_{1L}=0 \notag \\
&U^\textup{(j)}_1 A^\textup{(j)}_{11L}+U^\textup{(j+1)}_1 A^\textup{(j)}_{12L}+U^\textup{(j)}_2 A^\textup{(j)}_{13L}+U^\textup{(j+1)}_2 A^\textup{(j)}_{14L} + \epsilon_0 A^\textup{(j)}_{15L} + \phi A^\textup{(j)}_{16L} + E^\textup{(j)}_{2L}=0 \notag \\
&\forall j \in \left\{1,...n-1\right\} 
\end{align}
and the \eqref{eq.equilcond} in the form
\begin{align}
\label{eq.equilconf2}
&U^{(1)}_1 A_{1L}+U^{(1)}_2 A_{2L}+\epsilon_0 A_{3L}+ \phi A_{4L}+E_{3L}=0 \notag \\
&U^\textup{(n)}_1 A_{n1L}+U^\textup{(n)}_2 A_{n2L}+\epsilon_0 A_{n4L}+ \phi A_{n5L}+E_{nL}=0
\end{align}

in order to solve the algebraic system constituted by \eqref{eq.condeln3} and \eqref{eq.equilconf2}, it could be convenient to re-arrange the algebraic system following a matrix based procedure. Indeed, the known terms can be collected in the load vector $\textbf{L}$ with $2n+2$ dimension
\[
\textbf{L}=\left[
\begin{array}{c}
0,
0,
...,
F_z,
\mathcal{M}
\end{array}
\right]^T
\]
while the unknown parameters can be collected in the $\textbf{X}$ $(2n+2)$-vector as follows:
\[
\textbf{X}=\left[
\begin{array}{c}
U^{(1)}_1,
U^{(1)}_2,
U^{(2)}_1,
U^{(2)}_2,
...,
U^\textup{(i)}_1,
U^\textup{(i)}_2,
...,
U^\textup{(n)}_1,
U^\textup{(n)}_2,
\epsilon_0,
\phi
\end{array}
\right]^T
\]
so that the set of equations \eqref{eq.condeln3} and \eqref{eq.equilconf2} becomes
\begin{equation}
\label{eq.nmat}
\mathbb{P} \cdot \textbf{X}=\textbf{L}
\end{equation}
where $\mathbb{P}$ is a $[(2n+2) \times (2n+2)]$ square matrix containing the coefficients $A^\textup{(j)}_{iL}$, $E^\textup{(j)}_{iL}$, which are functions of both geometrical and constitutive parameters of the phases.\\
Finally, being the system \eqref{eq.nmat} of linear and algebraic type, provided that $\det\mathbb{P}\neq 0$\footnote{The possibility to invert the matrix $\mathbb{P}$ is ensured by invoking the uniqueness of the linear elastic solution, due to Kirchhoff’s theorem. This could appear not immediately evident if one directly tries to see the actual form of $\mathbb{P}$. However an analytical proof that the algebraic problem is well-posed is given by utilizing the \textsl{Mathematica}\textsuperscript\textregistered code, where the command $RowReduce$ is employed. This command performs a version of Gaussian elimination, adding multiples of rows together so as to produce zero elements when possible. The final matrix is in reduced row echelon form. If it is a non-degenerate square matrix, as in our case, $RowReduce[\mathbb{P}]$ gives the $IdentityMatrix[Length[\mathbb{P}]]$.}, it is thus possible to write the solution as follows:
\begin{equation}
\label{eq.nsol}
\textbf{X}=\mathbb{P}^{-1}\textbf{L}=\frac{adj[\mathbb{P}]}{\det\mathbb{P}}\textbf{L}=\frac{\widetilde{\mathbb{P}}}{\det\mathbb{P}}\textbf{L}, \qquad X_m=\frac{1}{\det\mathbb{P}} \sum^{m=2n+3}_{h=1} \widetilde{P}_{h/m}L_h
\end{equation}
where $adj[\mathbb{P}]=\widetilde{\mathbb{P}}$ is the adjoint matrix of $\mathbb{P}$ and then the Cramer rule has been employed.\\
Dually, in a linear elastostatic problem with displacement-prescribed boundary conditions, the number of unknowns reduces to $2n$. In such a case, the equations \eqref{eq.zcond} must be substituted with \eqref{eq.dispcondel} with $\epsilon_0$ and $\phi$ imposed for all the phases from \eqref{eq.epsfiel}, the interface conditions and the equilibria at the internal and external cylindrical boundaries remaining unchanged.
%\begin{align}
%\label{eq.equilcondisp}
%&\sigma_{rr}^{(1)}|_{r=R_0}=0 \\
%&u_r^\textup{(n)}|_{r=R_n}=0 \notag
%\end{align}
Recasting the \eqref{eq.condeln2} with $\epsilon$ and $\phi$ known in the form
\begin{align}
\label{eq.condeld3}
&U^\textup{(j)}_1 A^\textup{(j)}_{5D}+U^\textup{(j+1)}_1 A^\textup{(j)}_{6D}+U^\textup{(j)}_2 A^\textup{(j)}_{7D}+U^\textup{(j+1)}_2 A^\textup{(j)}_{8D} + E^\textup{(j)}_{1D}=\epsilon_0 A^\textup{(j)}_{9D} + \phi A^\textup{(j)}_{10D}\notag \\
&U^\textup{(j)}_1 A^\textup{(j)}_{11D}+U^\textup{(j+1)}_1 A^\textup{(j)}_{12D}+U^\textup{(j)}_2 A^\textup{(j)}_{13D}+U^\textup{(j+1)}_2 A^\textup{(j)}_{14D} + E^\textup{(j)}_{2D}=\epsilon_0 A^\textup{(j)}_{15D} + \phi A^\textup{(j)}_{16D} \notag \\
&\forall j \in \left\{1,...n-1\right\} 
\end{align}
and the \eqref{eq.equilcond} in the form
\begin{align}
\label{eq.equilcond2}
&U^{(1)}_1 A_{1D}+U^{(1)}_2 A_{2D}+E_{3D}=\epsilon_0 A_{3D}+ \phi A_{4D} \notag \\
&U^\textup{(n)}_1 A_{n1D}+U^\textup{(n)}_2 A_{n2D}+E_{nD}=\epsilon_0 A_{n4D}+ \phi A_{n5D}
\end{align}
in order to solve the algebraic system constituted by \eqref{eq.condeld3} and \eqref{eq.equilcond2}, it is again convenient to re-arrange the algebraic system following a matrix based procedure. Indeed the known terms can be collected in the displacement $2n$-vector \textbf{D}
\[
\textbf{D}=\left[
\begin{array}{cc}
A_{3D} & A_{4D}\\
A^{(2)}_{9D} & A^{(2)}_{10D}\\
A^{(2)}_{15D} & A^{(2)}_{16D}\\
... & ...\\
A^\textup{(i)}_{9D} & A^\textup{(i)}_{10D}\\
A^\textup{(i)}_{15D} & A^\textup{(i)}_{16D}\\
... & ...\\
A_{n4D} & A_{n5D}
\end{array}
\right] \cdot \left[
\begin{array}{c}
\epsilon_0 \\
\phi
\end{array}
\right]
\]
while the unknown parameters can be collected in the $2n$-dimensional vector \textbf{Y} as follows
\[
\textbf{Y}=\left[
\begin{array}{c}
U^{(1)}_1,
U^{(1)}_2,
U^{(2)}_1,
U^{(2)}_2,
...,
U^\textup{(i)}_1,
U^\textup{(i)}_2,
...,
U^\textup{(n)}_1,
U^\textup{(n)}_2
\end{array}
\right]^T
\]
so that the set of equations \eqref{eq.condeld3} and \eqref{eq.equilcond2} becomes
\begin{equation}
\label{eq.nmatd}
\mathbb{Q} \cdot \textbf{Y}=\textbf{D}
\end{equation}
where $\mathbb{Q}$ is a $[2n \times 2n]$ square matrix containing the new coefficients $A^\textup{(j)}_{iD}$, $E^\textup{(j)}_{iD}$, which are functions of both radii and elastic moduli of the phases as well.\\
Finally, being the system \eqref{eq.nmatd} of linear and algebraic type, provided that $\det\mathbb{Q}\neq 0$, it is possible to write the solution as follows:
\begin{equation}
\label{eq.nsold}
\textbf{Y}=\mathbb{Q}^{-1}\textbf{D}=\frac{adj[\mathbb{Q}]}{\det\mathbb{Q}}\textbf{D}=\frac{\widetilde{\mathbb{Q}}}{\det\mathbb{Q}}\textbf{D}, \qquad Y_m=\frac{1}{\det\mathbb{Q}} \sum^{m=2n+3}_{h=1} \widetilde{Q}_{h/m}D_h
\end{equation}
where $adj[\mathbb{Q}]=\widetilde{\mathbb{Q}}$ is the adjoint matrix of $\mathbb{Q}$ and then the Cramer rule has been employed.

\subsection*{Exact solutions for a 2-phase osteon model}
Focusing on a two-phase system and referring to the field solutions \eqref{eq.fieldurel}, \eqref{eq.fielduf2el} and \eqref{eq.fielduz2el}, the unknown constants result:
\begin{align*}
&V_0^{(1)} & &V_0^{(2)} & &V_3^{(1)} & &V_3^{(2)} \\
&U_1^{(1)} & &U_1^{(2)} & &U_2^{(1)} & &U_2^{(2)} 
\end{align*}
 
These can be found by imposing proper boundary and interface relations, which, under displacement-prescribed conditions, with an assigned contraction/dilation in the $z$-axis direction of the cylinder and null circumferential displacements at the bases, are:
\begin{align}
\label{eq.condel}
&\sigma_{rr}^{(1)}|_{r=R_0}=0 & & & &\notag\\
&\sigma_{rr}^{(1)}|_{r=R_1}=\sigma_{rr}^{(2)}|_{r=R_1} & &\sigma_{r\varphi}^{(1)}|_{r=R_1}=\sigma_{r\varphi}^{(2)}|_{r=R_1} &  &\sigma_{rz}^{(1)}|_{r=R_1}=\sigma_{rz}^{(2)}|_{r=R_1} \notag \\
&u_r^{(1)}|_{r=R_1}=u_r^{(2)}|_{r=R_1} & &u_\varphi^{(1)}|_{r=R_1}=u_\varphi^{(2)}|_{r=R_1} &  &u_z^{(1)}|_{r=R_1}=u_z^{(2)}|_{r=R_1}  \\
&\sigma_{rr}^{(2)}|_{r=R_2}=0 & & & & \notag
\end{align}
Finally, substitution of the solutions into the equations \eqref{eq.fieldurel},\eqref{eq.fielduf2el},\eqref{eq.fielduz2el} leads to find the exact analytical form of the displacements' components in the two phases, i.e. $u_r^{(1)}$, $u_{\varphi}^{(1)}$, $u_z^{(1)}$, $u_r^{(2)}$, $u_{\varphi}^{(2)}$ and $u_z^{(2)}$, from which one can then derive the whole strain and stress fields.\\
Starting from the obtained complete closed-form solutions, approximated expressions have been derived by considering moderate angle mismatches and a sufficiently small ratio between the lamellae thickness and the interface radius. With these assumptions, by also excluding differences in Poisson ratios for the transversely isotropic fibrils in order to have at one disposal compact expressions, the difference between the averages of the hoop stresses in the two lamellae --given by formula \eqref{App_law_avg}-- and the interface stress jumps have been examined, the latter one being described for sufficiently small angle mismatches (say up to $\Delta\beta=\pm 15^{\circ}$) by means of the approximated function:
\begin{equation}\label{App_law_int}
\frac{\llbracket\sigma_{\varphi\varphi}\rrbracket}{E\,\epsilon_0}=\frac{N(\beta)\Delta\,\beta^2+N(\pi/2-\beta)\Delta\,\beta}{D_1(\beta)\Delta\,\beta^2+D_1(\pi/2-\beta)\Delta\,\beta+D_2(\beta)}
\end{equation}
where $\Delta\beta=(\gamma-1)\beta$ and
\begin{equation}
\resizebox{.93\textwidth}{!}{$
\begin{array}{ll}
N(\xi)=(1-\alpha ) \alpha  \left\{\left[2 (3 \alpha +5) \nu ^2+\alpha  (5 \alpha +11) \nu +5\alpha (\alpha -1)  \right] \cos (2 \xi )-\left(\alpha  \nu +\alpha -2\nu ^2\right) \left[4(1+\hat{\delta}(\xi-\beta)) (\alpha +2 \nu +1) \cos (4 \xi )-(1+2\hat{\delta}(\xi-\beta)) (\alpha -1) \cos (6 \xi )\right]\right\}\\
D_1(\xi)=2 (1-\alpha) \left(\alpha -\nu ^2\right) \left[2(1+\hat{\delta}(\xi-\beta)) \left(\alpha  \nu +\alpha -2 \nu ^2\right) \cos (4 \xi )-4 \alpha  (\nu +1) \cos (2 \xi )\right]\\   
D_2(\xi)=2 \left(\alpha -\nu ^2\right) \left\{(\alpha -1) \left[\left(\alpha  \nu +\alpha -2 \nu ^2\right) \cos (4 \xi )-4 \alpha  (\nu +1) \cos (2 \xi )\right]+2 (\alpha -1) \nu ^2+\alpha  (3 \alpha +5) (1+\nu)\right\}
\end{array}
$}
\end{equation}
with $\hat{\delta}(\cdot)$ indicating the Kronecker delta symbol. A first order approximation of relation \eqref{App_law_int} for smaller angles mismatches and $\alpha$ near to 1 gives that 
\begin{equation}
\frac{\llbracket\sigma_{\varphi\varphi}\rrbracket}{E\,\epsilon_0}\approx\frac{(\alpha -1) \Delta \beta  \sin (2 \beta ) ((\nu -1) (2 \nu +1) \cos (2 \beta )+\nu )}{\nu ^2-1}
\end{equation} 

\section*{Helical to cylindrical coordinate system transformation}
\label{AppendixB}
Let us consider the helical coordinate system $\left\{r,t,c\right\}$ characterized by unit vectors\\
$\left\{\textbf{e}_r,\textbf{e}_t,\textbf{e}_c\right\}$. This coordinate system has unit vector $\textbf{e}_t$ tangent to helix; the unit vector $\textbf{e}_r$ is perpendicular to $\textbf{e}_t$ and the unit vector $\textbf{e}_c$ is perpendicular to the plane defined by $r$ and $t$.\\
Then, let us consider a new coordinate system characterized by the cylindrical system $\{r,\varphi,z\}$ with the unit vectors of such system $\left\{\textbf{e}_r,\textbf{e}_\varphi,\textbf{e}_z\right\}$. If the cylindrical system has the same origin of the helical one, the unit vector $\textbf{e}_r$ is coincident in the two coordinate systems. The angle between two unit vectors $\textbf{e}_t$ and $\textbf{e}_\varphi$ is identified by $\vartheta$, that is the same angle between two unit vectors $\textbf{e}_c$ and $\textbf{e}_z$.\\
Euler's theorem on the representation of rigid body rotations has many forms. The theorem concerns the characterization of a three-dimensional rotation by an angle $\vartheta$ about a specific axis, here indicated by the unit vector $\textbf{p}$. This theorem is represented by the formula:
\begin{equation}
\textbf{Q}=\textbf{I}+\textbf{P}\sin{\vartheta}+(1-\cos{\vartheta})\textbf{P}^2
\label{eq.p}
\end{equation}
where the three-dimensional skew-symmetric tensor $\textbf{P}$ with components $P_{ij}$ is introduced to represent the unit vector $\textbf{p}$,
\begin{equation}
\textbf{P}=
\left[
\begin{array}{ccc}
 0 & -p_3 & p_2 \\
 p_3 & 0 & -p_1 \\
 -p_2 & p_1 & 0 
\end{array}
\right] \quad \text{or} \quad P_{ij}=e_{ijk}p_k
\label{eq.pmatrix}
\end{equation}
It is easy to show that $\textbf{P}$ has the following properties:
\begin{equation}
\textbf{P}=-\textbf{P}^T, \quad \textbf{P}\textbf{p}=0, \quad \textbf{P}^2=\textbf{p}\otimes\textbf{p}-\textbf{I}, \quad \textbf{P}^3=-\textbf{P}
\label{eq.pprop}
\end{equation}
The transformation law between cylindrical and helical coordinate systems is characterized by a rotation about the radius. In this case the vector $\textbf{p}=\left\{1,0,0\right\}$ then the skew-symmetric tensor $\textbf{P}$ becomes:
\begin{equation}
\textbf{P}=
\left[
\begin{array}{ccc}
 0 & 0 & 0 \\
 0 & 0 & -1 \\
 0 & 1 & 0 
\end{array}
\right] \quad \text{or} \quad P_{ij}=e_{ijk}p_k
\label{eq.pmatrix2}
\end{equation}
By applying the formula \eqref{eq.p} the rotation matrix $\textbf{Q}$ can be obtained:
\begin{equation}
\textbf{Q}=
\left[
\begin{array}{ccc}
 1 & 0 & 0 \\
 0 & \cos{\vartheta} & -\sin{\vartheta} \\
 0 & \sin{\vartheta} & \cos{\vartheta} 
\end{array}
\right]
\label{eq.qmatrix}
\end{equation}
Then the transformation law between two coordinate systems is:
\begin{equation}
\left[
\begin{array}{ccc}
 e_r \\
 e_\varphi  \\
 e_z 
\end{array}
\right]
=
\left[
\begin{array}{ccc}
 1 & 0 & 0 \\
 0 & \cos{\vartheta} & -\sin{\vartheta} \\
 0 & \sin{\vartheta} & \cos{\vartheta} 
\end{array}
\right] \cdot
\left[
\begin{array}{ccc}
 e_r \\
 e_t  \\
 e_c 
\end{array}
\right]=
\textbf{Q} \cdot
\left[
\begin{array}{ccc}
 e_r \\
 e_t  \\
 e_c 
\end{array}
\right]
\label{eq.translaw}
\end{equation}
In a six-dimension space, the representation of a three-dimensional rotation by an angle $\vartheta$ about a specific axis, characterized by the unit vector $\textbf{p}$, is represented as a six-dimensional orthogonal tensor by the formula:
\begin{equation}
\hat{\textbf{Q}}=\textbf{I}+\hat{\textbf{P}}sin{\vartheta}+(1-\cos{\vartheta})\hat{\textbf{P}}^2+\tfrac{1}{3}\sin{\vartheta}(1-\cos{\vartheta})(\hat{\textbf{P}}+\hat{\textbf{P}}^3)+\tfrac{1}{6}(1-\cos{\vartheta})^2(\hat{\textbf{P}}^2+\hat{\textbf{P}}^4)
\label{eq.psixdim}
\end{equation}
where the six-dimensional skew-symmetric tensor $\hat{\textbf{P}}$ with components:
\begin{equation}
\hat{\textbf{P}}=\left[
\begin{array}{cccccc}
 0 & 0 & 0 & 0 & \sqrt{2}p_2 & -\sqrt{2}p_3 \\
 0 & 0 & 0 & -\sqrt{2}p_1 & 0 & \sqrt{2}p_3 \\
 0 & 0 & 0 & \sqrt{2}p_1 & \sqrt{2}p_2 & 0 \\
 0 & \sqrt{2}p_1 & -\sqrt{2}p_1 & 0 & p_3 & p_2 \\
 -\sqrt{2}p_2 & 0 & \sqrt{2}p_2 & -p_3 & 0 & p_1 \\
 \sqrt{2}p_3 & -\sqrt{2}p_3 & 0 & p_2 & -p_1 & 0
\end{array}
\right]
\label{eq.pmatrixsix}
\end{equation}
satisfies the following conditions:
\begin{equation}
\hat{\textbf{P}}=-\hat{\textbf{P}}^T, \quad \hat{\textbf{P}}^5+5\hat{\textbf{P}}^3+4\hat{\textbf{P}}=0, \quad \hat{\textbf{P}}^6+5\hat{\textbf{P}}^4+4\hat{\textbf{P}}^2=0
\label{eq.ppropsixdim}
\end{equation}
Matrices of six-dimension tensor components, according to Cowin's notation, are distinguished with the symbol "$\hat{}$". Note that the formulas \eqref{eq.p} and \eqref{eq.psixdim} show that a change in the orientation of $\textbf{p}$ is the same as a reversal of the direction of the angle from $\textbf{p},\vartheta$ to $-\textbf{p},-\vartheta$.\\
Formula \eqref{eq.psixdim} is of interest in anisotropic elasticity because the elasticity tensor can be expressed as a second rank tensor in six dimensions \cite{eigentensor}, as well as in its more traditional representation as a fourth rank tensor in three dimensions. Formula \eqref{eq.psixdim} connects the geometric operation in three dimensions to the matrix algebra of six dimensions. Since the tensor transformation rules for a second rank tensor rather than a fourth rank tensor apply, transformations of the reference coordinate system for the elasticity tensor may be accomplished in a very straightforward fashion using matrix multiplication.\\ 
The anisotropic form of Hooke's law is often written as $\sigma_{ij}=C_{ijkm}\epsilon_{km}$, where the $C_{ijkm}$ are the components of the three-dimensional fourth rank elasticity tensor.
There are three important symmetric restrictions on the fourth rank tensor components $C_{ijkm}$. These restrictions, which require that components with the subscripts $ijkm$ and $kmij$ be equal, follow from the symmetry of the stress tensor, the symmetry of the strain tensor, and the requirement that no work can be produced by the elastic material in a close loading cycle, respectively. Written as a linear transformation in six dimensions, Hooke's law has the representation $\boldsymbol{T}=\textbf{C}\cdot\boldsymbol{E}$ or 
\begin{equation}
\left[
\begin{array}{c}
 \sigma _{11} \\
 \sigma _{22} \\
 \sigma _{33} \\
 \sigma _{23} \\
 \sigma _{13} \\
 \sigma _{12}
\end{array}
\right]=
\left[
\begin{array}{cccccc}
 C_{11} & C_{12} & C_{13} & C_{14} & C_{15} & C_{16} \\
 C_{12} & C_{22} & C_{23} & C_{24} & C_{25} & C_{26} \\
 C_{13} & C_{23} & C_{33} & C_{34} & C_{35} & C_{36} \\
 C_{14} & C_{24} & C_{34} & C_{44} & C_{45} & C_{46} \\
 C_{15} & C_{25} & C_{35} & C_{45} & C_{55} & C_{56} \\
 C_{16} & C_{26} & C_{36} & C_{46} & C_{56} & C_{66}
\end{array}
\right]
\cdot
\left[
\begin{array}{c}
 \epsilon_{11} \\
 \epsilon_{22} \\
 \epsilon_{33} \\
 2\epsilon_{23} \\
 2\epsilon_{13} \\
 2\epsilon_{12}
\end{array}
\right]
\label{eq.costimatr}
\end{equation}
In the Voigt notation the components of $\textbf{C}$ and $C_{ijkm}$ are related by replacing the six-dimensional indexes $1, 2, 3, 4, 5$ and $6$ by the pairs of the three-dimensional indexes $1, 2$ and $3$; thus $1,2,3,4,5$ and $6$ become $11, 22, 33, 23$ or $32, 13$ or $31, 12$ or $21$, respectively. The members of the paired indexes $23$ or $32, 13$ or $31, 12$ or $21$ are equivalent because of the symmetry of the tensors of the stress and strain. The matrix $\textbf{C}$ in equation \eqref{eq.costimatr} is not a matrix of tensor components in six dimensions, although it is formed from the components of a three-dimensional fourth rank tensor.\\
Six-dimensional vector base and notations are introduced so that stress and strain can be considered as vectors in a six-dimensional vector space as well as second order rank tensors in three-dimensional Cartesian reference systems. The six-dimensional quantities will be indicated by the hat notation; thus, the six-dimensional vectors of stress and strain will be denoted by $\hat{\textbf{T}}$ and $\hat{\textbf{E}}$, respectively, whereas the three-dimensional second rank tensors of stress and strain are denoted by $\textbf{T}$ and $\textbf{E}$, respectively. The direct relationship between the components of $\hat{\textbf{T}}$ and $\textbf{T}$, and $\hat{\textbf{E}}$ and $\textbf{E}$, are dual representations given by
\begin{equation}
\hat{\textbf{T}}=
\left[
\begin{array}{c}
 \hat{\sigma} _{1} \\
 \hat{\sigma} _{2} \\
 \hat{\sigma} _{3} \\
 \hat{\sigma} _{4} \\
 \hat{\sigma} _{5} \\
 \hat{\sigma} _{6}
\end{array}
\right]
=\left[
\begin{array}{c}
 \sigma _{11} \\
 \sigma _{22} \\
 \sigma _{33} \\
\sqrt{2}\sigma _{23} \\
\sqrt{2}\sigma _{13} \\
\sqrt{2}\sigma _{12}
\end{array}
\right], \quad
\hat{\textbf{E}}=
\left[
\begin{array}{c}
 \hat{\epsilon} _{1} \\
 \hat{\epsilon} _{2} \\
 \hat{\epsilon} _{3} \\
 \hat{\epsilon} _{4} \\
 \hat{\epsilon} _{5} \\
 \hat{\epsilon} _{6}
\end{array}
\right]
=\left[
\begin{array}{c}
 \epsilon _{11} \\
 \epsilon _{22} \\
 \epsilon _{33} \\
\sqrt{2}\epsilon _{23} \\
\sqrt{2}\epsilon _{13} \\
\sqrt{2}\epsilon _{12}
\end{array}
\right]
\label{eq.relcow}
\end{equation}
where the shear components of these new six-dimensional stress and strain vectors are the shear components of these three-dimensional stress and strain tensors multiplied by $\sqrt{2}$. This $\sqrt{2}$ factor ensures that the scalar product of the two six-dimensional vectors is equal to the trace of the product of the corresponding second rank tensors, $\hat{\textbf{T}}\cdot\hat{\textbf{E}}=tr(\textbf{T}\cdot\textbf{E})$.\\
Introducing the Cowin's notation of equation \eqref{eq.relcow} into equation \eqref{eq.costimatr}, \eqref{eq.costimatr} can be rewritten in the form:
\begin{equation}
\hat{\textbf{T}}=\hat{\textbf{C}}\cdot\hat{\textbf{E}}
\label{eq.costimatrcow}
\end{equation}
which in explicit takes the form
\begin{equation}
\left[
\begin{array}{c}
 \sigma _{11} \\
 \sigma _{22} \\
 \sigma _{33} \\
 \sqrt{2}\sigma _{23} \\
 \sqrt{2}\sigma _{13} \\
 \sqrt{2}\sigma _{12}
\end{array}
\right]=
\left[
\begin{array}{cccccc}
 C_{11} & C_{12} & C_{13} & \sqrt{2}C_{14} & \sqrt{2}C_{15} & \sqrt{2}C_{16} \\
 C_{12} & C_{22} & C_{23} & \sqrt{2}C_{24} & \sqrt{2}C_{25} & \sqrt{2}C_{26} \\
 C_{13} & C_{23} & C_{33} & \sqrt{2}C_{34} & \sqrt{2}C_{35} & \sqrt{2}C_{36} \\
 \sqrt{2}C_{14} & \sqrt{2}C_{24} & \sqrt{2}C_{34} & C_{44} & C_{45} & C_{46} \\
 \sqrt{2}C_{15} & \sqrt{2}C_{25} & \sqrt{2}C_{35} & C_{45} & C_{55} & C_{56} \\
 \sqrt{2}C_{16} & \sqrt{2}C_{26} & \sqrt{2}C_{36} & C_{46} & C_{56} & C_{66}
\end{array}
\right]
\cdot
\left[
\begin{array}{c}
 \epsilon_{11} \\
 \epsilon_{22} \\
 \epsilon_{33} \\
 \sqrt{2}\epsilon_{23} \\
 \sqrt{2}\epsilon_{13} \\
 \sqrt{2}\epsilon_{12}
\end{array}
\right]
\label{eq.costimatrex}
\end{equation}
or
\begin{equation}
\left[
\begin{array}{c}
 \hat{\sigma} _{1} \\
 \hat{\sigma} _{2} \\
 \hat{\sigma} _{3} \\
 \hat{\sigma} _{4} \\
 \hat{\sigma} _{5} \\
 \hat{\sigma} _{6}
\end{array}
\right]=
\left[
\begin{array}{cccccc}
 \hat{C}_{11} & \hat{C}_{12} & \hat{C}_{13} & \hat{C}_{14} & \hat{C}_{15} & \hat{C}_{16} \\
 \hat{C}_{12} & \hat{C}_{22} & \hat{C}_{23} & \hat{C}_{24} & \hat{C}_{25} & \hat{C}_{26} \\
 \hat{C}_{13} & \hat{C}_{23} & \hat{C}_{33} & \hat{C}_{34} & \hat{C}_{35} & \hat{C}_{36} \\
 \hat{C}_{14} & \hat{C}_{24} & \hat{C}_{34} & \hat{C}_{44} & \hat{C}_{45} & \hat{C}_{46} \\
 \hat{C}_{15} & \hat{C}_{25} & \hat{C}_{35} & \hat{C}_{45} & \hat{C}_{55} & \hat{C}_{56} \\
 \hat{C}_{16} & \hat{C}_{26} & \hat{C}_{36} & \hat{C}_{46} & \hat{C}_{56} & \hat{C}_{66}
\end{array}
\right]
\cdot
\left[
\begin{array}{c}
 \hat{\epsilon}_{1} \\
 \hat{\epsilon}_{2} \\
 \hat{\epsilon}_{3} \\
 \hat{\epsilon}_{4} \\
 \hat{\epsilon}_{5} \\
 \hat{\epsilon}_{6}
\end{array}
\right]
\label{eq.costimatrex2}
\end{equation}
The relationship between the non-tensor Voigt notation $\textbf{C}$ and six-dimensional second rank tensor $\hat{\textbf{C}}$ components is easily constructed from equation \eqref{eq.costimatrex}; a table of this relationship is given in Mehrabadi and Cowin \cite{eigentensor}.
The symmetric matrix $\hat{\textbf{C}}$ can be shown to represent the components of a second rank tensor in a six-dimensional space, whereas the components of the matrix $\textbf{C}$ appearing in equation \eqref{eq.costimatr} do not form a tensor.
It is easy to prove that if the material has rhombic syngony in the helical system, then in the cylindrical coordinate system, the material has a monoclinic anisotropy. The monoclinic crystal system has exactly one pale of reflective symmetry. A material is said to have a plane of reflective symmetry with respect to a plane passing through the point. In particular, in the cylindrical coordinate the plane of elastic symmetry is $\varphi-z$. Remember that vector $\textbf{p}$ is equal to $\left\{1,0,0\right\}$, then, six-dimensional skew-symmetric tensor $\hat{\textbf{P}}$, becomes:
\begin{equation}
\hat{\textbf{P}}=\left[
\begin{array}{cccccc}
 0 & 0 & 0 & 0 & 0 & 0 \\
 0 & 0 & 0 & -\sqrt{2} & 0 & 0\\
 0 & 0 & 0 & \sqrt{2} & 0 & 0 \\
 0 & \sqrt{2} & -\sqrt{2} & 0 & 0 & 0 \\
 0 & 0 & 0 & 0 & 0 & 1 \\
 0 & 0 & 0 & 0 & -1 & 0
\end{array}
\right]
\label{eq.pmat}
\end{equation}
Then, from \eqref{eq.psixdim}, six-dimensional orthogonal tensor $\hat{\textbf{Q}}$ is equal to:
\begin{equation}
\hat{\textbf{Q}}=\left[
\begin{array}{cccccc}
 1 & 0 & 0 & 0 & 0 & 0 \\
 0 & \cos^2{\vartheta} & \sin^2{\vartheta} & -\tfrac{\sqrt{2}}{2}\sin{2\vartheta} & 0 & 0\\
 0 & \sin^2{\vartheta} & \cos^2{\vartheta} & \tfrac{\sqrt{2}}{2}\sin{2\vartheta} & 0 & 0 \\
 0 & \tfrac{\sqrt{2}}{2}\sin{2\vartheta} & -\tfrac{\sqrt{2}}{2}\sin{2\vartheta} & \cos{2\vartheta} & 0 & 0 \\
 0 & 0 & 0 & 0 & \cos{\vartheta} & \sin{\vartheta} \\
 0 & 0 & 0 & 0 & -\sin{\vartheta} & \cos{\vartheta}
\end{array}
\right]
\label{eq.qmat}
\end{equation}
In helical coordinate system the elastic matrix is:
\begin{equation}
\textbf{C}^{hel}=\left[
\begin{array}{cccccc}
 C^{hel}_{11} & C^{hel}_{12} & C^{hel}_{13} & 0 & 0 & 0 \\
 C^{hel}_{12} & C^{hel}_{22} & C^{hel}_{23} & 0 & 0 & 0 \\
 C^{hel}_{13} & C^{hel}_{23} & C^{hel}_{33} & 0 & 0 & 0 \\
 0 & 0 & 0 & 2C^{hel}_{44} & 0 & 0 \\
 0 & 0 & 0 & 0 & 2C^{hel}_{55} & 0 \\
 0 & 0 & 0 & 0 & 0 & 2C^{hel}_{66}
\end{array}
\right]
\label{eq.chel}
\end{equation}
where the vector stress is obtained multiplying the matrix $\textbf{C}^{hel}$ by the strain vector, in Voigt notation. The moduli of elasticity $C^{hel}_{ij}$ are linked to the elastic moduli:\\
$E_{rr}, E_{tt}, E_{cc}, \nu_{rt}, \nu_{tc}, \nu_{rc}, G_{rt}, G_{tc}, G_{rc}$.\\
By invoking the transformation law \eqref{eq.translaw} in six dimension, with reference to \eqref{eq.chel} and \eqref{eq.qmat}, the elastic matrix in cylindrical coordinate system becomes:
\begin{equation}
\textbf{C}^{cyl}=\left[
\begin{array}{cccccc}
 C^{cyl}_{11} & C^{cyl}_{12} & C^{cyl}_{13} & 2C^{cyl}_{14} & 0 & 0 \\
 C^{cyl}_{12} & C^{cyl}_{22} & C^{cyl}_{23} & 2C^{cyl}_{24} & 0 & 0 \\
 C^{cyl}_{13} & C^{cyl}_{23} & C^{cyl}_{33} & 2C^{cyl}_{34} & 0 & 0 \\
 C^{cyl}_{14} & C^{cyl}_{24} & C^{cyl}_{34} & 2C^{cyl}_{44} & 0 & 0 \\
 0 & 0 & 0 & 0 & 2C^{cyl}_{55} & C^{cyl}_{56} \\
 0 & 0 & 0 & 0 & C^{cyl}_{56} & 2C^{cyl}_{66}
\end{array}
\right]
\label{eq.ccyl}
\end{equation}
where the constants $C^{cyl}_{ij}$ are linked with the ones $C^{hel}_{ij}$ and the angle of the helix $\vartheta$ through the following relations:
\begin{align}
\label{eq.heltocyl}
&C^{cyl}_{11}=C^{hel}_{11}\notag \\
&C^{cyl}_{12}=C^{hel}_{12}\cos^2{\vartheta}+C^{hel}_{13}\sin^2{\vartheta}\notag \\
&C^{cyl}_{13}=C^{hel}_{13}\cos^2{\vartheta}+C^{hel}_{12}\sin^2{\vartheta}\notag\\
&C^{cyl}_{14}=(C^{hel}_{12}-C^{hel}_{13})\frac{\sin{2\vartheta}}{2}\notag\\
&C^{cyl}_{22}=C^{hel}_{22}\cos^4{\vartheta}+(C^{hel}_{23}+2C^{hel}_{44})\frac{\sin^2{2\vartheta}}{2}+C^{hel}_{33}\sin^4{\vartheta}\notag\\
&C^{cyl}_{23}=C^{hel}_{12}\cos^2{2\vartheta}+\frac{1}{4}(C^{hel}_{22}-2C^{hel}_{23}+C^{hel}_{33}-4C^{hel}_{44})\sin^2{2\vartheta}\notag\\
&C^{cyl}_{24}=\frac{1}{4}\Bigl[C^{hel}_{22}-C^{hel}_{33}+(C^{hel}_{22}-2C^{hel}_{23}+C^{hel}_{33}-4C^{hel}_{44})\cos{2\vartheta}\Bigr]\sin{2\vartheta}\\
&C^{cyl}_{33}=C^{hel}_{33}\cos^4{\vartheta}+(C^{hel}_{23}+2C^{hel}_{44})\frac{\sin^2{2\vartheta}}{2}+C^{hel}_{22}\sin^4{\vartheta}\notag\\
&C^{cyl}_{34}=\frac{1}{4}\Bigl[C^{hel}_{22}-C^{hel}_{33}+(4C^{hel}_{44}-C^{hel}_{33}+2C^{hel}_{23}-C^{hel}_{22})\cos{2\vartheta}\Bigr]\sin{2\vartheta}\notag\\
&C^{cyl}_{44}=C^{hel}_{44}\cos^2{2\vartheta}+\frac{1}{4}(C^{hel}_{22}-2C^{hel}_{23}+C^{hel}_{33})\sin^2{2\vartheta}\notag\\
&C^{cyl}_{55}=C^{hel}_{55}\cos^2{\vartheta}+C^{hel}_{66}\sin^2{\vartheta}\notag\\
&C^{cyl}_{56}=(C^{hel}_{66}-C^{hel}_{55})\frac{\sin{2\vartheta}}{2}\notag\\
&C^{cyl}_{66}=C^{hel}_{55}\sin^2{\vartheta}+C^{hel}_{66}\cos^2{\vartheta}\notag
\end{align}

%\newpage
%%%%%%%%%%%%%%%%%%%%%%%%%%%%%%%%%%%%%%%%%%%%%%%%%%%%%%%%%%%
\bibliographystyle{ieeetr}  
%\bibliography{OsteonRef}  %%% Remove comment to use the external .bib file (using bibtex).

\begin{thebibliography}{10}

\bibitem{Cowin1981}
S.~C. Cowin, {\em {Mechanical properties of bone}}.
\newblock American Society of Mechanical Engineers, 1981.

\bibitem{currey2006bones}
J.~D. Currey, {\em Bones: structure and mechanics}.
\newblock Princeton university press, 2006.

\bibitem{cowin1983m}
S.~C. Cowin, ``The mechanical and stress adaptive properties of bone,'' {\em
  Annals of Biomedical Engineering}, vol.~11, no.~3, pp.~263--295, 1983.

\bibitem{cowin1976}
S.~Cowin and D.~Hegedus, ``Bone remodeling {I}: theory of adaptive
  elasticity,'' {\em Journal of Elasticity}, vol.~6, no.~3, pp.~313--326, 1976.

\bibitem{hegedus1976}
D.~Hegedus and S.~Cowin, ``Bone remodeling {II}: small strain adaptive
  elasticity,'' {\em Journal of elasticity}, vol.~6, no.~4, pp.~337--352, 1976.

\bibitem{cowin2003ad}
S.~Cowin, ``Adaptive elasticity: A review and critique of a bone tissue
  adaptation model,'' {\em Engineering Transactions}, vol.~51, no.~2-3,
  pp.~113--193, 2003.

\bibitem{fratzl2007}
P.~Fratzl and R.~Weinkamer, ``Nature’s hierarchical materials,'' {\em
  Progress in materials Science}, vol.~52, no.~8, pp.~1263--1334, 2007.

\bibitem{zimmermann2015}
E.~A. Zimmermann and R.~O. Ritchie, ``Bone as a structural material,'' {\em
  Advanced healthcare materials}, vol.~4, no.~9, pp.~1287--1304, 2015.

\bibitem{gao2016}
Y.~A. Shin, S.~Yin, X.~Li, S.~Lee, S.~Moon, J.~Jeong, M.~Kwon, S.~J. Yoo, Y.-M.
  Kim, T.~Zhang, {\em et~al.}, ``Nanotwin-governed toughening mechanism in
  hierarchically structured biological materials,'' {\em Nature
  communications}, vol.~7, no.~1, pp.~1--10, 2016.

\bibitem{Launey2010}
M.~E. Launey, M.~J. Buehler, and R.~O. Ritchie, ``{On the Mechanistic Origins
  of Toughness in Bone},'' in {\em Annual Review of Materials Research},
  vol.~40, pp.~25--53, Annual Reviews, 2010.

\bibitem{Martin2015}
R.~B. Martin, D.~B. Burr, N.~A. Sharkey, and D.~P. Fyhrie, {\em {Skeletal
  Tissue Mechanics}}.
\newblock New York: Springer-Verlag, 2015.

\bibitem{Cowin2007}
S.~C. Cowin and S.~B. Doty, {\em {Tissue Mechanics}}.
\newblock Springer-Verlag New York, 2007.

\bibitem{zysset1999}
P.~K. Zysset, X.~E. Guo, C.~E. Hoffler, K.~E. Moore, and S.~A. Goldstein,
  ``Elastic modulus and hardness of cortical and trabecular bone lamellae
  measured by nanoindentation in the human femur,'' {\em Journal of
  biomechanics}, vol.~32, no.~10, pp.~1005--1012, 1999.

\bibitem{oftadeh2015biomechanics}
R.~Oftadeh, M.~Perez-Viloria, J.~C. Villa-Camacho, A.~Vaziri, and A.~Nazarian,
  ``Biomechanics and mechanobiology of trabecular bone: a review,'' {\em
  Journal of biomechanical engineering}, vol.~137, no.~1, 2015.

\bibitem{Cowin1986w}
S.~C. Cowin, ``{Wolff’s Law of Trabecular Architecture at Remodeling
  Equilibrium},'' {\em Journal of Biomechanical Engineering}, vol.~108,
  pp.~83--88, 02 1986.

\bibitem{cowin2012mixture}
S.~C. Cowin and L.~Cardoso, ``Mixture theory-based poroelasticity as a model of
  interstitial tissue growth,'' {\em Mechanics of Materials}, vol.~44,
  pp.~47--57, 2012.

\bibitem{Gupta2006a}
H.~Gupta, U.~Stachewicz, W.~Wagermaier, P.~Roschger, H.~Wagner, and P.~Fratzl,
  ``{Mechanical modulation at the lamellar level in osteonal bone},'' {\em
  Journal of Materials Research}, vol.~21, no.~08, pp.~1913--1921, 2006.

\bibitem{Weiner1999}
S.~Weiner, W.~Traub, and H.~D. Wagner, ``{Lamellar bone: structure– function
  relations.},'' {\em Journal of Structural Biology}, vol.~126, pp.~241--255,
  1999.

\bibitem{cowin2015flow}
S.~C. Cowin and L.~Cardoso, ``Blood and interstitial flow in the hierarchical
  pore space architecture of bone tissue,'' {\em Journal of biomechanics},
  vol.~48, no.~5, pp.~842--854, 2015.

\bibitem{cardoso2013ad}
L.~Cardoso, S.~P. Fritton, G.~Gailani, M.~Benalla, and S.~C. Cowin, ``Advances
  in assessment of bone porosity, permeability and interstitial fluid flow,''
  {\em Journal of biomechanics}, vol.~46, no.~2, pp.~253--265, 2013.

\bibitem{Ascenzi1968}
A.~Ascenzi and E.~Bonucci, ``{The compressive properties of single osteons.},''
  {\em The Anatomical record}, vol.~161, pp.~377--91, jul 1968.

\bibitem{Ascenzi2003}
M.-G. Ascenzi, A.~Ascenzi, A.~Benvenuti, M.~Burghammer, S.~Panzavolta, and
  A.~Bigi, ``{Structural differences between “dark” and “bright”
  isolated human osteonic lamellae},'' {\em Journal of Structural Biology},
  vol.~141, pp.~22--33, jan 2003.

\bibitem{Ascenzi2006}
M.~G. Ascenzi and A.~Lomovtsev, ``{Collagen orientation patterns in human
  secondary osteons, quantified in the radial direction by confocal
  microscopy},'' {\em Journal of Structural Biology}, vol.~153, no.~1,
  pp.~14--30, 2006.

\bibitem{Wagermaier2006a}
W.~Wagermaier, H.~S. Gupta, A.~Gourrier, M.~Burghammer, P.~Roschger, and
  P.~Fratzl, ``{Spiral twisting of fiber orientation inside bone lamellae},''
  {\em Biointerphases}, vol.~1, no.~1, p.~1, 2006.

\bibitem{Kazanci2006}
M.~Kazanci, P.~Roschger, E.~P. Paschalis, K.~Klaushofer, and P.~Fratzl, ``{Bone
  osteonal tissues by Raman spectral mapping: Orientation-composition},'' {\em
  Journal of Structural Biology}, vol.~156, no.~3, pp.~489--496, 2006.

\bibitem{Giraud-Guille1988}
M.~M. Giraud-Guille, ``{Twisted plywood architecture of collagen fibrils in
  human compact bone osteons},'' {\em Calcified Tissue International}, vol.~42,
  no.~3, pp.~167--180, 1988.

\bibitem{Schrof2014}
S.~Schrof, P.~Varga, L.~Galvis, K.~Raum, and A.~Masic, ``{3D Raman mapping of
  the collagen fibril orientation in human osteonal lamellae},'' {\em Journal
  of structural biology}, vol.~187, no.~3, pp.~266--275, 2014.

\bibitem{carnelli}
D.~Carnelli, P.~Vena, M.~Dao, C.~Ortiz, and R.~Contro, ``Orientation and
  size-dependent mechanical modulation within individual secondary osteons in
  cortical bone tissue,'' {\em Journal of The Royal Society Interface},
  vol.~10, no.~81, 2013.

\bibitem{martin-book}
R.~Bruce~Martin, D.~B. Burr, N.~A. Sharkey, and D.~P. Fyhrie, {\em Skeletal
  Tissue Mechanics}.
\newblock Springer, 2015.

\bibitem{Milovanovic}
P.~Milovnovic, A.~vom Scheidt, K.~Mletzko, G.~Sarau, K.~Puschel, M.~Djuric,
  M.~Amling, S.~Christiansen, and B.~Busse, ``Bone tissue aging affects
  mineralization of cement lines,'' {\em Bone}, vol.~110, 2018.

\bibitem{taber2020}
L.~A. Taber, {\em Continuum Modeling in Mechanobiology}.
\newblock Springer, 2020.

\bibitem{cowinmicrog}
S.~Cowin, ``On mechanosensation in bone under microgravity,'' {\em Bone},
  vol.~22, no.~5, pp.~119S--125S, 1998.

\bibitem{Taylor}
D.~Taylor, J.~G. Hazenberg, and T.~C. Lee, ``{Living with cracks: damage and
  repair in human bone.},'' {\em Nature materials}, vol.~6, no.~4,
  pp.~263--268, 2007.

\bibitem{Herman2010}
B.~C. Herman, L.~Cardoso, R.~J. Majeska, K.~J. Jepsen, and M.~B. Schaffler,
  ``{Activation of bone remodeling after fatigue: Differential response to
  linear microcracks and diffuse damage},'' {\em Bone}, vol.~47, no.~4,
  pp.~766--772, 2010.

\bibitem{cardoso2009ost}
L.~Cardoso, B.~C. Herman, O.~Verborgt, D.~Laudier, R.~J. Majeska, and M.~B.
  Schaffler, ``Osteocyte apoptosis controls activation of intracortical
  resorption in response to bone fatigue,'' {\em Journal of bone and mineral
  research}, vol.~24, no.~4, pp.~597--605, 2009.

\bibitem{klein2013osteocytes}
J.~Klein-Nulend, A.~D. Bakker, R.~G. Bacabac, A.~Vatsa, and S.~Weinbaum,
  ``Mechanosensation and transduction in osteocytes,'' {\em Bone}, vol.~54,
  no.~2, pp.~182--190, 2013.

\bibitem{Klein-Nulend}
J.~Klein-Nulend, R.~Bacabac, and M.~Mullender, ``Mechanobiology of bone
  tissue,'' {\em Pathologie biologie}, vol.~53, no.~10, pp.~576--580, 2005.

\bibitem{Ericksen}
E.~F. Eriksen, ``Cellular mechanisms of bone remodeling,'' {\em Reviews in
  Endocrine and Metabolic Disorders}, vol.~11, no.~4, pp.~219--227, 2010.

\bibitem{Taylor2}
D.~Taylor and T.~Lee, ``Microdamage and mechanical behaviour: predicting
  failure and remodelling in compact bone,'' {\em Journal of anatomy},
  vol.~203, no.~2, pp.~203--211, 2003.

\bibitem{lewis2017osteocyte}
K.~J. Lewis, D.~Frikha-Benayed, J.~Louie, S.~Stephen, D.~C. Spray, M.~M. Thi,
  Z.~Seref-Ferlengez, R.~J. Majeska, S.~Weinbaum, and M.~B. Schaffler,
  ``Osteocyte calcium signals encode strain magnitude and loading frequency in
  vivo,'' {\em Proceedings of the National Academy of Sciences}, vol.~114,
  no.~44, pp.~11775--11780, 2017.

\bibitem{CowinPNAS}
Y.~Han, S.~C. Cowin, M.~B. Schaffler, and S.~Weinbaum, ``Mechanotransduction
  and strain amplification in osteocyte cell processes,'' {\em Proceedings of
  the national academy of sciences}, vol.~101, no.~47, pp.~16689--16694, 2004.

\bibitem{Verbruggen2014}
S.~W. Verbruggen, T.~J. Vaughan, and L.~M. McNamara, ``{Fluid flow in the
  osteocyte mechanical environment: A fluid-structure interaction approach},''
  {\em Biomechanics and Modeling in Mechanobiology}, vol.~13, no.~1,
  pp.~85--97, 2014.

\bibitem{wu2013matrix}
D.~Wu, M.~B. Schaffler, S.~Weinbaum, and D.~C. Spray, ``Matrix-dependent
  adhesion mediates network responses to physiological stimulation of the
  osteocyte cell process,'' {\em Proceedings of the National Academy of
  Sciences}, vol.~110, no.~29, pp.~12096--12101, 2013.

\bibitem{wagermaier2015fragility}
W.~Wagermaier, K.~Klaushofer, and P.~Fratzl, ``Fragility of bone material
  controlled by internal interfaces,'' {\em Calcified tissue international},
  vol.~97, no.~3, pp.~201--212, 2015.

\bibitem{peterlik2006brittle}
H.~Peterlik, P.~Roschger, K.~Klaushofer, and P.~Fratzl, ``From brittle to
  ductile fracture of bone,'' {\em Nature materials}, vol.~5, no.~1,
  pp.~52--55, 2006.

\bibitem{Nalla2003}
R.~K. Nalla, J.~H. Kinney, and R.~O. Ritchie, ``{Mechanistic fracture criteria
  for the failure of human cortical bone.},'' {\em Nature materials}, vol.~2,
  no.~3, pp.~164--8, 2003.

\bibitem{Ritchie2021}
R.~O. Ritchie, ``Toughening materials: enhancing resistance to fracture,'' {\em
  Philosophical Transactions of the Royal Society A: Mathematical, Physical and
  Engineering Sciences}, vol.~379, p.~20200437, Aug. 2021.

\bibitem{Ritchie2011}
R.~O. Ritchie, ``{The conflicts between strength and toughness},'' {\em Nature
  Materials}, vol.~10, no.~11, pp.~817--822, 2011.

\bibitem{Moyle1984}
D.~D. Moyle and R.~W. Bowden, ``{Fracture of human femoral bone},'' {\em
  Journal of Biomechanics}, vol.~17, pp.~203--213, jan 1984.

\bibitem{Libonati}
F.~Libonati, G.~X. Gu, Z.~Qin, L.~Vergani, and M.~J. Buehler, ``Bone-inspired
  materials by design: toughness amplification observed using 3d printing and
  testing,'' {\em Advanced Engineering Materials}, vol.~18, no.~8,
  pp.~1354--1363, 2016.

\bibitem{Kendall1975}
K.~Kendall, ``{Control of cracks by interfaces in composites},'' {\em
  Proceedings of the Royal Society of London . Series A}, vol.~341,
  pp.~409--428, 1975.

\bibitem{ping2022mineralization}
H.~Ping, W.~Wagermaier, N.~Horbelt, E.~Scoppola, C.~Li, P.~Werner, Z.~Fu, and
  P.~Fratzl, ``Mineralization generates megapascal contractile stresses in
  collagen fibrils,'' {\em Science}, vol.~376, no.~6589, pp.~188--192, 2022.

\bibitem{Sabet}
F.~A. Sabet, A.~Raeisi~Najafi, E.~Hamed, and I.~Jasiuk, ``Modelling of bone
  fracture and strength at different length scales: a review,'' {\em Interface
  focus}, vol.~6, no.~1, p.~20150055, 2016.

\bibitem{Yeni2003}
Y.~N. Yeni, F.~J. Hou, T.~Ciarelli, D.~Vashishth, and D.~P. Fyhrie,
  ``{Trabecular shear stresses predict in vivo linear microcrack density but
  not diffuse damage in human vertebral cancellous bone},'' {\em Annals of
  Biomedical Engineering}, vol.~31, no.~6, pp.~726--732, 2003.

\bibitem{wolfram2016}
U.~Wolfram, J.~J. Schwiedrzik, M.~J. Mirzaali, A.~B{\"u}rki, P.~Varga,
  C.~Olivier, F.~Peyrin, and P.~K. Zysset, ``Characterizing microcrack
  orientation distribution functions in osteonal bone samples,'' {\em Journal
  of microscopy}, vol.~264, no.~3, pp.~268--281, 2016.

\bibitem{weinbaum1994}
S.~Weinbaum, S.~C. Cowin, and Y.~Zeng, ``A model for the excitation of
  osteocytes by mechanical loading-induced bone fluid shear stresses,'' {\em
  Journal of biomechanics}, vol.~27, no.~3, pp.~339--360, 1994.

\bibitem{cowin1998amp}
S.~C. Cowin and S.~Weinbaum, ``Strain amplification in the bone mechanosensory
  system,'' {\em The American journal of the medical sciences}, vol.~316,
  no.~3, pp.~184--188, 1998.

\bibitem{you2001}
L.~You, S.~C. Cowin, M.~B. Schaffler, and S.~Weinbaum, ``A model for strain
  amplification in the actin cytoskeleton of osteocytes due to fluid drag on
  pericellular matrix,'' {\em Journal of biomechanics}, vol.~34, no.~11,
  pp.~1375--1386, 2001.

\bibitem{ICTAM2021}
N.~Pugno, ``Bioinspired nanomechanics: a contribution for the centenary of the
  griffith's theory,'' in {\em Plenary opening lecture at 25th International
  Congress of Theoretical and Applied Mechanics}, Aug. 2021.

\bibitem{Lanyon1979}
L.~E. Lanyon and S.~Bourn, ``{The influence of mechanical function on the
  development and remodeling of the tibia. An experimental study in sheep.},''
  {\em The Journal of bone and joint surgery. American volume}, vol.~61,
  pp.~263--73, mar 1979.

\bibitem{Behiri1984}
J.~Behiri and W.~Bonfield, ``{Fracture mechanics of bone—The effects of
  density, specimen thickness and crack velocity on longitudinal fracture},''
  {\em Journal of Biomechanics}, vol.~17, no.~1, pp.~25--34, 1984.

\bibitem{nalla2005}
R.~K. Nalla, J.~J. Kruzic, J.~H. Kinney, and R.~O. Ritchie, ``Mechanistic
  aspects of fracture and r-curve behavior in human cortical bone,'' {\em
  Biomaterials}, vol.~26, no.~2, pp.~217--231, 2005.

\bibitem{ZIMMERMANN2014}
E.~A. Zimmermann, B.~Gludovatz, E.~Schaible, B.~Busse, and R.~O. Ritchie,
  ``Fracture resistance of human cortical bone across multiple length-scales at
  physiological strain rates,'' {\em Biomaterials}, vol.~35, no.~21,
  pp.~5472--5481, 2014.

\bibitem{Wang1997}
T.~L. Norman and Z.~Wang, ``Microdamage of human cortical bone: incidence and
  morphology in long bones,'' {\em Bone}, vol.~20, no.~4, pp.~375--379, 1997.

\bibitem{Chan2010}
K.~S. Chan, C.~K. Chan, and D.~P. Nicolella, ``{NIH Public Access},'' vol.~45,
  no.~3, pp.~427--434, 2010.

\bibitem{Krajcinovic1987}
Krajcinovic, ``{Constitutive laws for engineering materials: Theory and
  application},'' {\em Engineering Fracture Mechanics}, vol.~28, no.~3, p.~365,
  1987.

\bibitem{BruceMartin1982}
R.~{Bruce Martin} and D.~B. Burr, ``{A hypothetical mechanism for the
  stimulation of osteonal remodelling by fatigue damage},'' {\em Journal of
  Biomechanics}, vol.~15, pp.~137--139, jan 2017.

\bibitem{lakes1981}
J.~Yang and R.~Lakes, ``Transient study of couple stress effects in compact
  bone: torsion,'' 1981.

\bibitem{cowinporo}
S.~C. Cowin, ``Bone poroelasticity,'' {\em Journal of biomechanics}, vol.~32,
  no.~3, pp.~217--238, 1999.

\bibitem{cowin2011fabric}
S.~C. Cowin and L.~Cardoso, ``Fabric dependence of wave propagation in
  anisotropic porous media,'' {\em Biomechanics and modeling in
  mechanobiology}, vol.~10, no.~1, pp.~39--65, 2011.

\bibitem{cardoso2012role}
L.~Cardoso and S.~C. Cowin, ``Role of structural anisotropy of biological
  tissues in poroelastic wave propagation,'' {\em Mechanics of Materials},
  vol.~44, pp.~174--188, 2012.

\bibitem{Fraldi2007}
M.~Fraldi, L.~Nunziante, and F.~Carannante, ``{Axis-Symmetrical Solutions for n
  -Plies Functionally Graded Material Cylinders under Strain No-Decaying
  Conditions},'' {\em Mechanics of Advanced Materials and Structures}, vol.~14,
  no.~3, pp.~151--174, 2007.

\bibitem{Cutolo}
A.~Cutolo, A.~Carotenuto, F.~Carannante, N.~Pugno, and M.~Fraldi, ``Analytical
  solutions for monoclinic/trigonal structures replicating multi-wall carbon
  nano-tubes for applications in composites with elastomeric/polymeric
  matrix,'' in {\em High-performance elastomeric materials reinforced by
  nano-carbons}, pp.~193--234, Elsevier, 2020.

\bibitem{Fraldi2002}
M.~Fraldi and S.~C. Cowin, ``{Chirality in the Torsion of Cylinders with
  Trigonal Symmetry},'' {\em Journal of Elasticity}, vol.~69, no.~1-3,
  pp.~121--148, 2002.

\bibitem{cowin2013b}
S.~C. Cowin, {\em Continuum mechanics of anisotropic materials}.
\newblock Springer Science \& Business Media, 2013.

\bibitem{cowin1987}
S.~C. Cowin and M.~M. Mehrabadi, ``On the identification of material symmetry
  for anisotropic elastic materials,'' {\em The Quarterly Journal of Mechanics
  and Applied Mathematics}, vol.~40, no.~4, pp.~451--476, 1987.

\bibitem{zysset1995ft}
P.~Zysset and A.~Curnier, ``An alternative model for anisotropic elasticity
  based on fabric tensors,'' {\em Mechanics of Materials}, vol.~21, no.~4,
  pp.~243--250, 1995.

\bibitem{Vercher-Martnez2015}
A.~Vercher-Martnez, E.~Giner, C.~Arango, and F.~{Javier Fuenmayor},
  ``{Influence of the mineral staggering on the elastic properties of the
  mineralized collagen fibril in lamellar bone},'' {\em Journal of the
  Mechanical Behavior of Biomedical Materials}, vol.~42, pp.~243--256, 2015.

\bibitem{yoon2008}
Y.~J. Yoon and S.~C. Cowin, ``The estimated elastic constants for a single bone
  osteonal lamella,'' {\em Biomechanics and modeling in mechanobiology},
  vol.~7, no.~1, pp.~1--11, 2008.

\bibitem{math}
I.~Wolfram~Research, {\em Mathematica}.
\newblock Wolfram Research, Inc., 2015.

\bibitem{fraldi2014sef}
M.~Fraldi, ``The paradox of the element carved upside down in the neiko gate
  that cannot be straightened,'' 2014.

\bibitem{Ascenzi1999}
M.~G. Ascenzi, ``{A first estimation of prestress in so-called circularly
  fibered osteonic lamellae},'' {\em Journal of Biomechanics}, vol.~32, no.~9,
  pp.~935--942, 1999.

\bibitem{Ansys}
I.~ANSYS, {\em ANSYS Mechanical User's Guide}.
\newblock 2013.

\bibitem{RaeisiNajafi2007}
A.~{Raeisi Najafi}, A.~R. Arshi, M.~R. Eslami, S.~Fariborz, and M.~H.
  Moeinzadeh, ``{Micromechanics fracture in osteonal cortical bone: A study of
  the interactions between microcrack propagation, microstructure and the
  material properties},'' {\em Journal of Biomechanics}, vol.~40, no.~12,
  pp.~2788--2795, 2007.

\bibitem{Abdel-wahab2012}
A.~A. Abdel-wahab, A.~R. Maligno, and V.~V. Silberschmidt, ``{Micro-scale
  modelling of bovine cortical bone fracture : Analysis of crack propagation
  and microstructure using X-FEM},'' {\em Computational Materials Science},
  vol.~52, no.~1, pp.~128--135, 2012.

\bibitem{Ural2013}
A.~Ural and S.~Mischinski, ``{Multiscale modeling of bone fracture using
  cohesive finite elements},'' {\em Engineering Fracture Mechanics}, vol.~103,
  pp.~141--152, 2013.

\bibitem{Giner2014}
E.~Giner, C.~Arango, A.~Vercher, and F.~J. Fuenmayor, ``{Numerical modelling of
  the mechanical behaviour of an osteon with microcracks},'' {\em Journal of
  the Mechanical Behavior of Biomedical Materials}, vol.~37, pp.~109--124,
  2014.

\bibitem{cornetti2006}
P.~Cornetti, N.~Pugno, A.~Carpinteri, and D.~Taylor, ``Finite fracture
  mechanics: a coupled stress and energy failure criterion,'' {\em Engineering
  Fracture Mechanics}, vol.~73, no.~14, pp.~2021--2033, 2006.

\bibitem{taylor2005fracture}
D.~Taylor, P.~Cornetti, and N.~Pugno, ``The fracture mechanics of finite crack
  extension,'' {\em Engineering fracture mechanics}, vol.~72, no.~7,
  pp.~1021--1038, 2005.

\bibitem{pugno2006}
N.~Pugno, M.~Ciavarella, P.~Cornetti, and A.~Carpinteri, ``A generalized
  paris’ law for fatigue crack growth,'' {\em Journal of the Mechanics and
  Physics of Solids}, vol.~54, no.~7, pp.~1333--1349, 2006.

\bibitem{Caler1989}
W.~E. Caler and D.~R. Carter, ``{Bone creep-fatigue damage accumulation},''
  {\em Journal of Biomechanics}, vol.~22, pp.~625--635, jan 1989.

\bibitem{Carter1977}
D.~Carter and W.~Hayes, ``{Compact bone fatigue damage—I. Residual strength
  and stiffness},'' {\em Journal of Biomechanics}, vol.~10, pp.~325--337, jan
  1977.

\bibitem{Griffin1997}
L.~V. Griffin, J.~C. Gibeling, R.~B. Martin, V.~A. Gibson, and S.~M. Stover,
  ``{Model of flexural fatigue damage accumulation for cortical bone},'' {\em
  Journal of Orthopaedic Research}, vol.~15, no.~4, pp.~607--614, 1997.

\bibitem{Pattin1996}
C.~A. Pattin, W.~E. Caler, and D.~R. Carter, ``{Cyclic mechanical property
  degradation during fatigue loading of cortical bone},'' {\em Journal of
  Biomechanics}, vol.~29, no.~1, pp.~69--79, 1996.

\bibitem{wang2022mechanical}
L.~Wang, X.~You, L.~Zhang, C.~Zhang, and W.~Zou, ``Mechanical regulation of
  bone remodeling,'' {\em Bone Research}, vol.~10, no.~1, pp.~1--15, 2022.

\bibitem{cardoso2015changes}
L.~Cardoso and M.~B. Schaffler, ``Changes of elastic constants and anisotropy
  patterns in trabecular bone during disuse-induced bone loss assessed by
  poroelastic ultrasound,'' {\em Journal of Biomechanical Engineering},
  vol.~137, no.~1, p.~011008, 2015.

\end{thebibliography}

\end{document}